\documentstyle[seceq,preprint,epsf]{ptptex}

\newcommand{\bfig}{\begin{figure}}
\newcommand{\efig}{\end{figure}}

\def\one-loop{\mbox{\scriptsize one-loop}}

\def\a{\alpha}

\def\s{\sigma}
\def\th{\theta}
\def\thb{\bar{\theta}}

\def\psivector{\mbox{\boldmath$\psi_{Q}$}}
\def\Upsilonvector{\mbox{\boldmath$\Upsilon$}}
\def\Qvector{\mbox{\boldmath$Q$}}
\def\Omegavector{\mbox{\boldmath$\Omega$}}
\def\Pivector{\mbox{\boldmath$\Pi$}}
\def\Zvector{\mbox{\boldmath$Z$}}
\def\dg{\dagger}

\def\ud{\underline}

\def\G{\Gamma}

\def\e{\epsilon}
\def\eb{\bar{\epsilon}}

\newskip\humongous \humongous=0pt plus 1000pt minus 1000pt

\newif\ifdtup

\def\ket#1{\left| #1\right\rangle}

\def\beq{\begin{equation}}
\def\eeq{\end{equation}}

\def\beqn{\begin{eqnarray}}
\def\eeqn{\end{eqnarray}}
\relax

\def\G2{{\; \rm GeV/}c^2}
\def\G{\; \rm GeV}

\def\dotx{\dotx{\dot\overline{x}}}

\relax

\title{
 $USp(2k)$ Matrix Model \footnote{ based on the lecture given by
  the first author at the 13th Nishinomiya-Yukawa Memorial
 Symposium ``Dynamics of Fields and Strings'' (November 12-13,1998)
  and on the lecture by the second author at the YITP workshop
 (November 16-18, 1998).}
}

\author{
Hiroshi {\sc Itoyama}\footnote{ itoyama@funpth.phys.sci.osaka-u.ac.jp} 
and Asato {\sc Tsuchiya}\footnote{
tsuchiya@funpth.phys.sci.osaka-u.ac.jp }
}

\inst{
        Department of Physics,\\
        Graduate School of Science, Osaka University,\\
        Toyonaka, Osaka 560-0043, Japan}

\recdate{
}

\abst{
    We review  the construction  and theoretical implications of  the
 $USp(2k)$ matrix model in zero dimension introduced in ref.~\cite{IT1,IT2}. 
  It is argued that  the model provides a constructive approach
  to  $Type I$ superstrings and is at the same time dynamical  theory
 of spacetime points.
  Three subjects are discussed : semiclassical pictures and series of
  degenerate perturbative vacua associated with the worldvolume
 representation of the model,  the formation of extended ($D-$)
objects from the fermionic integrations via the (non-)abelian
  Berry phase, and the Schwinger-Dyson/loop equations which exhibit
 the joining-splitting interactions required.
}

\begin{document}

\nopagebreak

\maketitle

\section{Introduction}

   In this symposium, there are several talks devoted to the recent 
developments of the large $N$ limit of gauge theories and the
 matrix models for string unification.  We will discuss
  a specific matrix model in zero dimension based on $USp(2k)$ 
  introduced in ref.~\cite{IT1,IT2}.  ( We will refer to  zero
  dimensional matrix model in general as reduced matrix model.)
 We will present the criteria, the logic and the construction leading to the
 model as well as theoretical implications from our present understanding.
 
  Gauge fields and strings - the two notions occupying our mind- have had
 interesting relationship: which of the two notions is more fundamental has
   shifted from one to the other over the decades.
 Without making our talk historical, let us begin with mentioning that
 our current practise is to construct  string theory from  noncommuting
 matrix degrees of freedom  which originated from gauge fields.
 The major goal is 
  to overcome  the difficulties of the first
 quantized superstring theory  which have prevented us from
   predicting  physical quantities:  this will include 
 the one associated with the existence of
 the infinitely degenerate perturbative vacua.
  Let us first recall the five consistent perturbative superstrings in ten
  dimensions constructed by the end of 1984.

\bigskip
 
{\tiny
\renewcommand{\arraystretch}{2}
\begin{tabular}[t]{|c|c|} \hline
10 dim  N =2 (32 supercharges) & 10 dim  N=1 (16 supercharges)\\  \hline
type IIA \hspace{1cm} type IIB & 
type I \hspace{0.5cm} $SO(32)$ het \hspace{0.5cm}  $E_8 \times E_8$ het
\\ \hline
\end{tabular}
}

\bigskip

\noindent
  Type  $IIB$ superstrings  are related  to type $I$ superstrings by the
 twist operation and  the addition of the open-string sectors. 
    The rest  are related by the Wilson lines and the T duality in nine
 dimensions and by the S duality.

 The reduced model of type $IIB$ superstrings  has been proposed before
 \cite{IKKT}. We will focus on the  reduced model
  which  descends from the first quantized  nonorientable
 type $I$ superstrings
 \cite{GSano}: they are related to heterotic strings
   by S duality.
The $USp(2k)$ matrix model thus has a phenomenological
  perspective accessible to us by the presence of gauge bosons,
 matter fermions and other properties.

 The reduced model in general lays its basis on the 
 correspondence with
 the covariant Green-Schwarz superstrings
 in the Schild gauge \cite{Schild}.
  In this sense, the applicability of the reduced model is by no means 
  limited to low energy phenomena although its equivalence 
 with the first quantized  critical superstrings has so far been
  eatablished  at the level of classical  equations of motion on the
   two-dimensional worldsheet.
  One dimensional matrix model \cite{BFSS}
 of $M$ theory \cite{M} has, on the other hand,
 obtained  successes  on the agreement of the spectrum and other properties
 with the low energy eleven-dimensional
 supergravity theory.

  It has been demonstrated in \cite{IT1,IT2} that the $USp(2k)$ matrix model
 is uniquely selected by the three requirements:  
 
\bigskip

{\bf Requirements :}

 1)having eight dynamical and eight
 kinematical supercharges.

 2)obtained by an appropriate projection
 from the $IIB$ matrix model and an addition of the degrees of freedom
 corresponding to open strings.

 3)nonorientable.

\bigskip

  In the next section,  we summarize the criteria and the logic leading to
   the model \cite{IT2}.
  We will  begin with presenting the
 closed string sector of the model.   This will include
   introduction of the $USp$ projector and its commutativity with
  $8+8$ supersymmetry.  We also discuss
  the reduced model- Green-Schwarz  correspondence \cite{IKKT}. 
  After  the discussion \cite{Itsu}
   on  the open string sector, the loop variables and the Chan-Paton symmetry,
  we will present  the action of the model in its final
 form.

  The reduced matrix model is a constructive approach to superstring theory.
    It is at the same time dynamical theory of spacetime points, which
   we  briefly discuss in the last subsection.
  In quantum mechanics, space is a dynamical variable while time is
 considered to be a parameter. In (relativistic) quantum field theory,
  both space and time are parameters. In reduced model, both space and time
  are  dynamical variables appearing as eigenvalue distributions.
    This is an ideal setup for pursuing
   quantum gravity  which regards spacetime as a derived concept.

 In the subsequent sections, we
 discuss three subjects which are relevant to the properties
  of the $USp(2k)$ matrix model.
 Leaving aside the issue of lifting the degenerate perturbative
 vacua and the true scaling limit of the model,  the matrix model
  permits us  to consider a series of such vacua
  with $D$ objects through its $T$ dualized worldvolume representation. 
 We will discuss \cite{IT2} in section three a particular series of
 perturbative vacua
 associated with the $USp(2k)$ matrix model and the consistency of the model
 with the literature, examining properties of $4d,5d,6d$  worldvolume
  field theories. 
 In section four, we study the model by $T$-dualizing in the time direction,
 namely, by the $T$-dualized quantum mechanics.
  The main purpose here is to reveal the existence of branes as quantized
 degrees of freedom which the degrees of freedom in the fundamental
 representation are responsible for. This is done by examining
 properties of the fermionic integrations via the (non)-abelian
 Berry phase \cite{IM,CIK}. 
  We find formation of extended objects such as
  Dirac monopoles and its nonabelian generalization.
 
  The complete consideration of the model can be established by
 the Schwinger-Dyson/loop equations \cite{Itsu}, which are considered to be
 the second quantized formulation of the reduced matrix model. 
      We present the derivation in the final section.
 These loop equations
 exhibit a complete set of the joining and splitting interactions
 required for the nonorientable  $Type I$ superstrings. The study
 at the linearized level  provides us with the Virasoro conditions and
 the mixed Dirichlet/Neumann boundary conditions acting on the
 closed and open loop variables.
 
  We will make remarks not mentioned in  the original papers along
 the way the discussion goes.   Unless it is necessary, 
  we will suppress the $USp$ indices  in most of our discussion.
  More extensive references are found in the original papers
  \cite{IT1,IT2,Itsu,IM,CIK}.

\section{Criteria and logic leading to the model}

\subsection{projection from $IIB$ matrix model}
  We begin with the closed string sector. According
  to the criteria mentioned in the introduction, this sector should be
 obtained from the action of the  $IIB$ matrix model via the appropriate
 projection which  we will determine in the next subsection.
\subsubsection{$IIB$ matrix model}
    The action of the $IIB$ matrix  model  is
\beqn
   S_{IIB}(\ud{v}_{M},  \ud{\Psi} )  =
  \frac{1}{g^{2}} Tr \left( \frac{1}{4} \left[\ud{v}_{M},
 \ud{v}_{N} \right]
\left[ \ud{v}^{M}, \ud{v}^{N} \right] - \frac{1}{2}
\bar{\ud{\Psi}} \Gamma^{M} \left[ \ud{v}_{M},
\ud{\Psi} \right] \right) \;\;.
\eeqn
The symbols with underlines  lie in the adjoint representation
 of $U(2k)$ and
\beqn
\label{eq:phiv}
\Psi  =  \left(
 \lambda , 0, \psi_{\Phi_{1}}, 0 ,\psi_{\Phi_{2}}, 0  ,  \psi_{\Phi_{3}}, 0 ,
  0, \bar{\lambda}, 0,
 \bar{\psi}_{\Phi_{1}}, 0,
 \bar{\psi}_{\Phi_{2}},  0, \bar{\psi}_{\Phi_{3}} \right)^t \;\;.
\eeqn
 is a thirty two component Majorana-Weyl spinor
satisfying
\beqn
  C \bar{\Psi}^t = \Psi  \;
 , \; \Gamma_{11} \Psi =  \Psi \;\;\;.
\label{eq:majorana-weyl con}
\eeqn
  Later we will also use
\beqn
\Phi_{i} = \frac{1}{\sqrt{2}} \left( v_{3+i} +i v_{{6+i}}
 \right) \;\;.  \; 
\eeqn
 The ten dimensional gamma matrices have been denoted by 
 $\Gamma^{M}$.

\subsubsection{covariant Green-Schwarz superstrings in the Schild gauge}
   In the large $k$ limit, one can check that  this action
$S_{IIB}(\ud{v}_{M},  \ud{\Psi} )$ goes to that of
  the covariant Green-Schwarz superstrings in the Schild gauge. 
   Let us sketch how this is seen. In the large $k$ limit,  the group $SU(\infty)$
   goes to the group of area preserving diffeomorphisms  $(APD)$ on, for
 example, torus. ( We ignore the issue of the worldsheet geometry to consider
     and other subtleties here.)  The generators  are represented by
  two index objects $L_{\vec{m}}$, $\vec{m} = (m_{1}, m_{2})^{t}$
   with  $m_{1}$, $m_{2}$ being integers.  The algebra reads
\beqn
\left[ L_{\vec{m}}, L_{\vec{n}} \right] =  \left( m_{1}n_{2}-
  m_{2}n_{1} \right) L_{\vec{m}+\vec{n}}  \;\;.
\eeqn
  The $\ud{v}_{M}$ are expanded in $L_{\vec{m}}$   and can be written as
\beqn
  \ud{v}_{M}  &=&  \int \frac{d^{2}\sigma}{(2\pi)^{2}}  \tilde{\ud{v}}_{M}
  \left( \vec{\sigma} \right) \tilde{L} \left( \vec{\sigma} \right) 
   \nonumber \\
   &\equiv&   \tilde{L}_{ \tilde{\ud{v}}_{M}}  \;\;,
\eeqn
  where
\beqn
  \tilde{L} \left( \vec{\sigma} \right)   \equiv \sum_{\vec{m}}
   e^{-i\vec{m} \vec{\sigma}} L_{\vec{m}}  \;\;.
\eeqn 
  Using
\beqn
  \left[ \tilde{L}_{f},  \tilde{L}_{g} \right] =
 - \tilde{L}_{ \{ f,g \}_{P.B} }  \;\;,
\eeqn
  one can readily  derive
\beqn
  \lim_{k \rightarrow \infty} S_{IIB}  &=&  S_{Schild}  \\
 S_{Schild}  &\equiv& \frac{1}{2\pi \alpha^{\prime}}
  \int   \frac{d^{2}\sigma}{(2\pi)^{2}}  \left( \frac{1}{2}
   \tilde{\Sigma}_{MN} \tilde{\Sigma}^{MN} + 2i \tilde{\bar{\psi}} \Gamma^{M}
 \left\{  \tilde{\ud{v}}_{M}, \tilde{\psi} \right\}_{P.B} \right) \;\;.
\eeqn
  Here
\beqn
  \tilde{\Sigma}_{MN} \equiv \left\{ \tilde{\ud{v}}_{M},  \tilde{\ud{v}}_{M}
  \right\}_{P.B.}  \;\;.
\eeqn

    On the other hand,  starting from  the covariant  Green-Schwarz action,
  we can fix the local $\kappa$ symmetry  via the condition \cite{IKKT}
\beqn
  \theta^{1} = \theta^{2} \equiv \psi  \;\;,
\eeqn
  where $\theta^{1,2}$  are Majorana-Weyl spinor fields on the worldsheet
 having the sama chirality.
  We obtain
\beqn
 S_{GS}^{(fix)} &=& - \frac{1}{2\pi \alpha^{\prime}}  \int  
 \frac{d^{2}\sigma}{(2\pi)^{2}}   \left( \sqrt{-\frac{1}{2} \Sigma^{2}}
  \right.  \nonumber \\
 &&   \left. +2i \bar{\psi}\Gamma^{M}\left\{X_{M}, \psi \right\}_{P.B.}
  \right) \;\;
\eeqn
  in the Nambu-Goto form.
  Here   $ \Sigma_{MN} \equiv \left\{ X_{M},  X_{N}
  \right\}_{P.B.} $.
  Equation of motion  obtained from $ S_{GS}^{(fix)}$ reduces to
  that from $S_{Schild}$ provided  $\partial_{a} \left( \Sigma^{2}
  \right) = 0 $, which  can be  shown  by using again equation of motion
  obtained from $S_{Schild}$.
  
\subsubsection{USp/So projector}
  In order to make the closed string sector nonorientable, we need
     a projection of  $u(2k)$  Lie algebra  valued matrices,
  which corresponds to the twist operation $\Omega$ on the worldsheet.
 Natural structure to consider   is an embedding of
 $usp$ and $so$ Lie algebras  into the  $u(2k)$  Lie algebra.
  In both cases,   it is expedient to introduce the following projector:
\beqn
\label{projector}
 \hat{\rho}_{\mp} \bullet  \equiv \frac{1}{2} \left( \bullet \mp F^{-1}
 \bullet^{t}  F \right) \;\;
\eeqn
 Using this projector, one can uniquely decompose $u(2k)$ Lie algebra
 valued matrices  into the adjoint(= symmetric) and the antisymmetric
 representations of the $usp$ Lie algebra or into the adjoint(= antisymmetic)
  and symmetric  representations of the $so$ Lie algebra.  This is
  schematically drawn as

\bigskip

\begin{tabular}{lcc}
$F = \left(
\begin{array}{cc}
0 & I \\
-I & 0
\end{array}
\right)$ &
$U(2k)$ \mbox{adj} &
$\begin{array}{l}
\nearrow \, USp \ \mbox{adj(= sym)} \\
\searrow \, USp \ \mbox{asym}
\end{array}$ \\
$F = \left(
\begin{array}{cc}
0 & I \\
I & 0
\end{array}
\right)$ &
$U(2k)$ \mbox{adj} &
$\begin{array}{l}
\nearrow \, SO \ \mbox{adj(= asym)} \\
\searrow \, SO \ \mbox{sym}
\end{array}$
\end{tabular}

\bigskip

  We have found out in  references \cite{IT2,Itsu}  that the analysis
 based on the planar diagrams,  the consistency with the worldvolume
  field theory and the Chan-Paton factor of the open loop variable
 all lead to the choice of the $USp$ case.  In this talk we will only
 include this third discussion in subsection $2.3$.

 Let $L_{\vec{m}}$ be a generator belonging to the antisymmetric
 representation of $USp$.  In this case,
  $L_{\vec{m}}^{t} =L_{\vec{m}^{t}}$ with
 $\vec{m}^{t} \equiv \left( -m_{1},m_{2} \right)$.   We find
\beqn
\label{twistop}
 F^{-1}   \ud{v}_{M} F = \ud{v}_{M}^{t}  = 
 \int \frac{d^{2}\sigma}{(2\pi)^{2}}  \tilde{\ud{v}}_{M} 
 \left(-\sigma_{1},\sigma_{2} \right) \tilde{L} \left(
 \vec{\sigma} \right)  \;\;.
\eeqn
  The matrix $F$  is in fact the matrix counterpart of
  the twist operation $\Omega$.  If
 $L_{\vec{m}}$ be a generator belonging to the symmetric(=adjoint)
 representation of $USp$, we obtain an extra minus sign 
 to eq. (\ref{twistop}), telling us an orientifold operation.
 
\subsubsection{reduced model of closed nonorientable superstrings}

To summarize, the closed string sector of the reduced matrix model 
descending from the type $I$ superstrings  must take the following form:
\beqn
 S_{close}  \equiv   S_{0} ( v_{m} , \Phi_{I},
 \lambda, \psi_{\Phi_{I}},
\bar{\Phi}_{I}, \bar{\lambda},  \bar{\psi}_{\Phi_{I} })
 = S_{IIB}( \hat{\rho}_{b\mp}\ud{v}_{M},  \hat{\rho}_{f\mp}\ud{\Psi}  ) \;\;.
\label{eq:equiv}
\eeqn
  Here  $\hat{\rho}_{b\mp}$ is a diagonal matrix acting on  the vector
  indices while $\hat{\rho}_{f\mp}$ is a diagonal matrix
  acting on the spinor indices.  Each entry of these two matrices is
  either $\hat{\rho}_{-}$ or $\hat{\rho}_{+}$.
How these are chosen  while preserving  $8+8$  supersymmetry is the subject of
 the next subsection.


\subsection{$USp$ projector and supersymmetry}
  In order to make $8+8$ supersymmetry and the $USp$ projector compatible,
    the model must implement a set of conditions under which  the projectors
  $\hat{\rho}_{b \mp}$, $\hat{\rho}_{f \mp}$,
 and dynamical $\delta^{(1)}$ as well as
 kinematical $\delta^{(2)}$ supersymmetry commute.
This constraint of commutativity turns out to be very stringent
 and essentially leads to the unique possibility.  We will repeat the
 discussion  of \cite{IT2}.
  Start with
\beqn
   \delta^{(1)} \ud{v}_{M} &=& i \bar{\epsilon} \Gamma_{M} \ud{\Psi}
\label{eq:d11} \\
\delta^{(1)} \ud{\Psi} &=& \frac{i}{2} \left[ \ud{v}_{M}, \ud{v}_{N} \right]
 \Gamma^{MN} \epsilon  \label{eq:d12} \\
 \delta^{(2)}\ud{v}_{M} &=& 0 \label{eq:d21} \\
\delta^{(2)} \ud{\Psi} &=& \xi \;\;\;.  \label{eq:d22}
\eeqn
 Let us write generically
\beqn
\label{eq:prv}
 v_{M} &\equiv&  \delta_{M}^{~N} \hat{\rho}_{b \mp}^{(N)}
 \ud{v}_{N} \;\;\;  \nonumber \\
 \Psi_{A} &\equiv&  \delta_{AB} \hat{\rho}_{f \mp}^{(B)} \ud{\Psi}_B \;\;.
\eeqn
  The condition
 $\left[\hat{\rho}_{b \mp}, \delta^{(1)} \right] \ud{v}_{M} =0$  together
 with eq.~(\ref{eq:d11}) gives
\beqn
\label{eq:1}
 \sum_{A=1}^{32} \left( \bar{\epsilon} \Gamma_{M}\right)_{A} \left(
\hat{\rho}_{f \mp}^{(A)}-  \hat{\rho}_{b \mp}^{(M)} \right)  \ud{\Psi}_{A}
= 0 \;\;,
\eeqn
  with index $M$ not summed.
  The condition
\beqn
\label{eq:con2}
\left.
\left[\hat{\rho}_{f \mp}, \delta^{(1)} \right] \ud{\Psi}
\right|_{\ud{v}_{M}
\rightarrow \hat{\rho}_{b \mp}\ud{v}_{M} } =0
\eeqn
 together with eq.~(\ref{eq:d12}) provides
\beqn
\label{eq:2}
 \left( 1 - \hat{\rho}_{f \mp}^{(A)} \right) \left[ \hat{\rho}_{b \mp}^{(M)}
 \ud{v}_{M},  \hat{\rho}_{b \mp}^{(N)} \ud{v}_{N} \right]
  \left( \Gamma^{MN} \epsilon \right)_{A} =0  \;\;\;.
\eeqn
  The restriction  at eq.~(\ref{eq:con2}) comes from the fact that
 eq.~(\ref{eq:d12}) is true only on shell.
Eq. ~(\ref{eq:d21})  does not give us anything new  while
$\left[\hat{\rho}_{f \mp}, \delta^{(2)} \right] \ud{\Psi} =0$ with
eq.~(\ref{eq:d22}) gives
\beqn
 \xi_{A} 1 = \xi_{A} \hat{\rho}_{f \mp}^{(A)} 1 \;\;\;,
\label{eq: condition kin susy}
\eeqn
  with index $A$ not summed.

  In order to proceed further, we rewrite  eq.~(\ref{eq:prv}) explicitly as
\beqn
 \hat{\rho}_{b \mp}^{(M)}  &\equiv& \Theta (M\in {\cal M}_{-}) \hat{\rho}_{-}
 +  \Theta (M\in {\cal M}_{+}) \hat{\rho}_{+}  \nonumber \\
\hat{\rho}_{f \mp}^{(A)} &\equiv&  \Theta (A\in {\cal A}_{-}) \hat{\rho}_{-}
+  \Theta (A\in {\cal A}_{+}) \hat{\rho}_{+} \;\;\; ,
\label{eq: explicit proj}
\eeqn
where
\beq
 {\cal M}_{-} \cup {\cal M}_{+}
 =  \{ \{ \; 0,1,2,3,4,5,6,7,8, 9 \; \} \} \;\; ,
\;\;
  {\cal M}_{-} \cap {\cal M}_{+} = \phi \;\;,
\label{eq: M condition}
\eeq
\beq
 {\cal A}_{-} \cup {\cal A}_{+}
 = \{ \{ \; 1,2,5,6,9,10,13,14,19,20,23,24,27,28,31,32 \; \} \} \;\; ,
\;\;
  {\cal A}_{-}\cap  {\cal A}_{+}=\phi \; .
\label{eq: A condition}
\eeq
  We find that  eq.~(\ref{eq:1}) gives
\beqn
\label{eq:set1}
  \left( \bar{\epsilon} \Gamma_{M_{-}} \right)_{A_{+}} = \left( \bar{\epsilon}
 \Gamma_{M_{+}} \right)_{A_{-}} =0 \;\;,
\eeqn
     while eq.~(\ref{eq:2}) gives
\beqn
\label{eq:set2}
  \left( \Gamma^{M_{-}N_{+}} \epsilon \right)_{A_{-}} &=& 0 \nonumber \\
 \left( \Gamma^{M_{-}N_{-}} \epsilon \right)_{A_{+}}  &=&
\left( \Gamma^{M_{+}N_{+}} \epsilon  \right)_{A_{+}} =0 \;\;.
\eeqn
Equation (\ref{eq: condition kin susy}) gives
\beqn
\label{eq:am}
\xi_{A_{-}} = 0 \;\;.
\eeqn
 As  we consider the case of eight kinematical
supersymmetries, the number of elements of the sets
  denoted by $\mbox{\boldmath $\sharp$} ({\cal A}_{\pm})$  must be
\beq
\mbox{\boldmath $\sharp$} ({\cal A_{-}}) = 8 \;\; \mbox{ and } \;\;  
\mbox{\boldmath $\sharp$}({\cal A_{+}}) = 8
\label{eq:number of A+A-} \;.
\eeq

  Eqs.~(\ref{eq:set1}) and (\ref{eq:set2})  are regarded as the ones which
determine the anticommuting parameter $\e$, and
the sets  ${\cal A}_{+}$, ${\cal A}_{-}$, ${\cal M}_{+}$ and
${\cal M}_{-}$. In addition they must satisfy the conditions 
(\ref{eq: M condition}), (\ref{eq: A condition}) and
(\ref{eq:number of A+A-}).
 
 We search for solutions by first trying out as an input an appropriate
 thirty-two component anticommuting parameter $\e$
satisfying Majorana-Weyl condition.
Given $\e$, we  see if  we can determine ${\cal A}_{+}$, ${\cal A}_{-}$,
 ${\cal M}_{+}$ and ${\cal M}_{-}$ successfully.

  We have tried out many cases.
The case leading to our model is
\beq
\e = (\e_0, 0, \e_1, 0, 0,0,0,0, 0, \eb_0, 0 , \eb_1, 0,0,0,0  )^t \;\; .
\eeq
Note that $\e_0$, $\e_1$, $\eb_0$ and $\eb_1$ are two-component
anticommuting parameters.

We have found out  that
\beqn
 \hat{\rho}_{b\mp}
 &=& diag
(\hat{\rho}_{-},\hat{\rho}_{-},\hat{\rho}_{-},
 \hat{\rho}_{-},\hat{\rho}_{-},\hat{\rho}_{+},
 \hat{\rho}_{+},\hat{\rho}_{-}, \hat{\rho}_{+},\hat{\rho}_{+} )
\nonumber \\
 \hat{\rho}_{f\mp}   &=&
\hat{\rho}_{-} 1_{(4)} \otimes
\left( \begin{array}{cccc}
	1_{(2)}& & & \\
	       & 0 & & \\
	       &   & 1_{(2)} & \\
	       &   &         &0
	\end{array}
\right)
+
\hat{\rho}_{+} 1_{(4)} \otimes
\left( \begin{array}{cccc}
	0& & & \\
	       & 1_{(2)} & & \\
	       &   & 0 & \\
	       &   &         & 1_{(2)}
	\end{array}
\right)
 \;\; ,
\eeqn
  is  a solution to the above set of equations.  We adopt this choice
  as the projectors of our model.
  
We have found  only one solution other than  this one, which is given
 by
\beq
\e = (\e_0, 0, \e_1, 0, 0,0,0,0, 0, 0,0,0,0, \eb_2, 0 , \eb_3   )^t \;\; .
\label{eq:80susy para}
\eeq
  The consistent sets
\beqn
{\cal A}_{-} &=&  \{ \{ \; 1,2,5,6,27,28,31,32 \; \} \} \;\; ,
\nonumber\\
{\cal A}_{+} &=&  \{ \{ \; 9,10,13,14,19,20,23,24 \; \} \} \;\; ,
\label{eq:A-A+08}
\eeqn
\beqn
{\cal M}_{-} &=& \{\{ \; 4,7 \;\}\}  \;\; ,
\nonumber\\
{\cal M}_{+} &=&   \{\{ \; 0,1,2,3,5,6,8,9 \;\}\} \;\;.
\label{eq:M-M+08}
\eeqn
have been obtained.
The projectors (\ref{eq: explicit proj}) are
\beqn
 \hat{\rho}_{b\mp}
 &=& diag
(\hat{\rho}_{+},\hat{\rho}_{+},\hat{\rho}_{+},
 \hat{\rho}_{+},\hat{\rho}_{-},\hat{\rho}_{+},
 \hat{\rho}_{+},\hat{\rho}_{-}, \hat{\rho}_{+},\hat{\rho}_{+} )
\nonumber \\
 \hat{\rho}_{f\mp}   &=&
\hat{\rho}_{-} 1_{(4)} \otimes
\left( \begin{array}{cccc}
	1_{(2)}& & & \\
	       & 0 & & \\
	       &   & 0 & \\
	       &   &         & 1_{(2)}
	\end{array}
\right)
+
\hat{\rho}_{+} 1_{(4)} \otimes
\left( \begin{array}{cccc}
	0& & & \\
	       & 1_{(2)} & & \\
	       &   & 1_{(2)} & \\
	       &   &         & 0
	\end{array}
\right)
 \;\; .
\label{eq:2-8 case projector}
\eeqn
  This is the case considered in ref. \cite{Mtwist,Ba}
 in the context  of $M$ theory compactification to
  the lightcone heterotic strings.
   The spinorial parameters $\e_0$, $\e_1$, $\eb_2$ and $\eb_3$ in
  eq. (\ref{eq:80susy para}) are all real, however,  and the  closed string sector
 obtained from this choice is not regarded as a projection from
  the  $IIB$ matrix model.

\subsection{closed and open loops and Chan-Paton symmetry}
      Loop variables play a decisive role in the second
 quantized  formulation of the theory;  they eventually act as string
 fields. This will be discussed in
 section $5$.  We include  the part of
 the discussion here to appreciate  the role played by the degrees of
 freedom in the (anti-)fundamental representation and
  the attendant Chan-Paton symmetry.

\subsubsection{adding fundamentals and flavor symmetry}
   It is well known  that nonorientable closed strings by themselves
 are not consistent.   We need to add degrees of freedom corresponding to
 open strings, keeping $8+8$  susy.  Let us, therefore, consider  the
  following ``hypermultiplet''  in the fundamental/antifundamental
 representation  of $usp(2k)$:
\beqn 
   \left( Q_{i}, \tilde{Q}^{i}, \psi_{Q}{}_{i}, \psi_{\tilde{Q}}^{i}
 \right) \;\;.
\eeqn
 The number of this multiplet is denoted as  $n_{f}$.

To make this ``flavour''symmetry (= local gauge symmetry of strings) manifest,
 let us introduce complex $2n_{f}$ dimensional vectors
\beqn
 {\bf Q} \equiv  \left\{ \begin{array}{ll}
       Q_{(f)}\;\;, &  f=1 \sim n_{f}  \\
     F^{-1} \tilde{Q}_{(f- n_{f})}\;\;, & f=n_{f} +1 \sim 2n_{f} \;,
   \end{array}
   \right.   \;\;\;
 {\bf Q}^{*} \equiv  \left\{ \begin{array}{ll}
       Q_{(f)}^{*} \;\;, &  f=1 \sim n_{f}  \\
      \tilde{Q}_{(f- n_{f})}^{\ast} F \;\;, & f=n_{f} +1 \sim 2n_{f}  \;.
   \end{array}
   \right.     
\eeqn
 Similarly,
\beqn
  \psivector   \equiv  \left\{ \begin{array}{ll}
      \psi_{ Q_{(f)}} \;\;, &  f=1 \sim n_{f}  \\
     F^{-1} \psi_{\tilde{Q}_{(f- n_{f})} } \;\;, & f=n_{f} +1 \sim 2n_{f} \;, 
   \end{array}
   \right.     
   \psivector^{\ast}    \equiv  \left\{ \begin{array}{ll}
       \overline{\psi}_{Q_{(f)} } \;\;, &  f=1 \sim n_{f}  \\
    \overline{\psi}_{\tilde{Q}_{(f- n_{f})} } F \;\;, & f=n_{f} +1
 \sim 2n_{f} \;.
   \end{array}
   \right.    
\eeqn
  We denote the $f$-th components of these vectors  by ${\bf Q}_{(f)}$
 etc.

\subsubsection{Choice of variables}

 Let us first introduce a discretized path-ordered exponential
 which represents a configuration of a string in momentum superspace:
\beqn
U[p^{M}_{.},\eta_{.};n_1,n_0] \equiv
P \exp (-i \sum_{n=n_0}^{n_1} (p^{M}_{n} v_{M}+\bar{\eta}_{n} \Psi))
= \stackrel{\leftarrow}{\prod_{n=n_0}^{n_1}} 
 \exp (-i p^{M}_{n} v_{M}-i \bar{\eta}_{n} \Psi)\;\;, \nonumber
\eeqn
  where  $p^{M}_{n}$ and $\eta_n$ are respectively the sources
 or the momentum distributions for $v_{M}$ and those for $\Psi$.
The closed loop is then defined by
\beq
\Phi[p^{M}_{.},\eta_{.};n_1,n_0] \equiv Tr U[p^{M}_{.},\eta_{.};n_1,n_0]\;\;.
\eeq
To consider an open loop, let us introduce
$\Xi = \left( \xi, \xi^{*} \right)$ as bosonic sources for $\Qvector_{(f)}$
 and $\Qvector_{(f)}^{\ast}$, and $\Theta = \left(\theta, \bar{\theta} \right)$
as  Grassmannian ones for $\psivector_{(f)}$ and $\psivector_{(f)}^{\ast}$:
$ \left( \Xi  \Omegavector_{(f)} \right)  = \xi \Qvector_{(f)}
                      +F^{-1} \xi^{\ast}\Qvector_{(f)}^{\ast}$,
 $\left( \Theta \Upsilonvector_{(f)} \right) = \theta \psivector_{(f)} 
+F^{-1} \bar{\theta} \psivector^{\ast}_{(f)}.$
   We write these collectively as
\beq
\left( \Lambda \Pivector_{(f)} \right) \equiv \left( \Xi \Omegavector_{(f)}
 \right)
 +
                      \left( \Theta  \Upsilonvector_{(f)}  \right)\;\;.
\eeq
  The open loop is defined by
\beq
\Psi_{f'f}[k^{m}_{.},\zeta;l_1,l_0;\Lambda',\Lambda]
\equiv   \left( \Lambda'  \Pivector_{(f')} \right)
F U[k^{m}_{.},\zeta_{.};l_1,l_0] 
\left( \Lambda  \Pivector_{(f)} \right) \;\;,
\eeq
  where $f$ and $f'$  are the Chan-Paton indices.
  The open and closed loops generate  all of the observables in the theory
under question.
 
  We now turn to the question of the nonorientability of the
 closed and the open loops. 
Using $v^{t}_{M}=\mp F v_{M} F^{-1}, \Psi^{t}= \mp F \Psi F^{-1},$
 and $F^{t}= -F$, we readily obtain
\beqn
\Phi[p^{M}_{.},\eta_{.};n_1,n_0] 
=Tr(\stackrel{\rightarrow}{\prod_{n=n_0}^{n_1}} 
\exp (-i p^{M}_{n} v^{t}_{M}-i \bar{\eta}_{n} \Psi^{t}))
=\Phi[\mp p^{M}_{.},\mp \eta_{.};n_0,n_1]\;\;,
\label{csnonori}
\eeqn
and 
\beq
\Psi_{f'f}[k^{M}_{.},\zeta_{.};l_1,l_0;\Lambda',\Lambda]
=-\Psi_{ff'}[\mp k^{M}_{.},\mp \zeta_{.};l_0,l_1;\Lambda,\Lambda']\;\;.
\label{osnonori}
\eeq
These equations relate a string configuration to the one with its
orientation and the Chan-Paton factor reversed and drawn pictorially
 as
\begin{figure}[t]
\epsfysize=3cm 
\centerline{\epsfbox{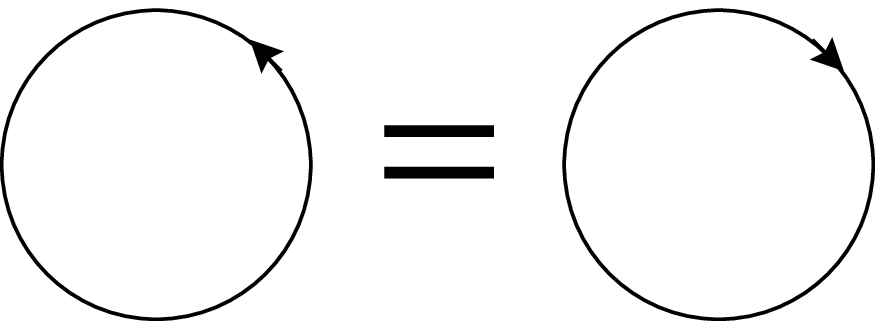}}
\label{fig1}
\end{figure}

\begin{figure}[t]
\epsfysize=3cm 
\centerline{\epsfbox{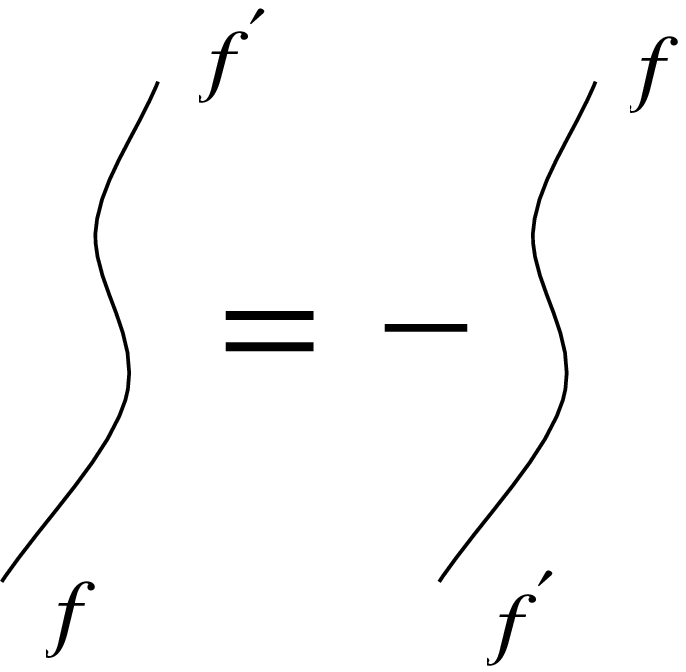}}
\label{fig2}
\end{figure}

  The minus signs in front of
 $p^{m}_{.}$, $\eta_{.}$, $k^{m}_{.}$ and $\zeta_{.}$ in eq.
 (\ref{csnonori}) and  eq. (\ref{osnonori}) reflect the orientifold
structure of the $USp(2k)$ matrix model.

The overall minus sign in the last line of (\ref{osnonori}) 
 is of interst: it comes from $F^{t}=-F$ of the $usp$ Lie algebra
 and implies the $SO(2 n_f)$ gauge group.  This is the cleanest one
  of the three rationales for the choice of the $usp$  Lie algebra:
 we present this as a table.

\bigskip

\begin{tabular}{ccccc}
         &             & & original(worldsheet) & \\
         & C-P factor  & & Lie algebra & \\
\hline
$-$ $\colon$ & $so(2n_{f})$ & $\Leftrightarrow$ & $usp(2k)$ &
 $\Leftarrow$ our choice \\
$+$ $\colon$ & $usp(2n_{f})$ & $\Leftrightarrow$ & $so(2k)$ &
\end{tabular}

\bigskip

  In order to inherit the infrared stability of perturbative vacua
  \cite{GS,IMo} of superstrings, the model must be based on
the $usp$ as opposed to the $so$ Lie algebra and  $n_f =16$.
This latter property also follows from the anomaly cancellation
of the $6D$ worldvolume gauge theory \cite{IT2}, which we will
 discuss in subsection $3.2$.

\subsection{the model}
 
\subsubsection{the action of the $USp$ matrix model}
  We finally  come to the action of our $USp(2k)$ reduced matrix model.
  It is obtained from the dimensional reduction of
 ${\cal N}=2, d=4$ $USp(2k)$ supersymmetric gauge theory with one
hypermultiplet in the antisymmetric representation and $n_{f}$ hypermultiplets
in the fundamental representation. This makes manifest the presence of the
eight dynamical supercharges.
 In the ${\cal N}=1$ superfield notation  with
  spacetime dependence all dropped, we have a vector superfield $V$ and a
  chiral superfield $\Phi \equiv \Phi_{1}$ which are $usp$ Lie algebra valued
\beqn
\label{eq:adj}
  V^{t} = - FV F^{-1} \;\;, \;\; \Phi^{t} = - F \Phi F^{-1}  \;\;,
 \;\; V^{\dagger} = V \;\;, \;\;  {\rm with} \;\;  F =
    \left(
 \begin{array}{cc}
   0 & I  \\
   -I & 0  
  \end{array}   \right)\;\;,
\eeqn
 and the two chiral superfields $\Phi_{I} ,\; I =2,3$ in the antisymmetric
 representation which obey
\beqn
\label{eq:asym}
  \Phi_{I}^{t}=  F \Phi_{I} F^{-1} \;\;\; {\rm for\; I=2,3}\;\;.
\eeqn 
 We can write
$V =  \hat{\rho}_{-} \ud{V},\;\;   \Phi_{1} = \hat{\rho}_{-}
 \ud{\Phi_{1}} ,\;\;
   \Phi_{I}  =   \hat{\rho}_{+} \ud{\Phi_{I}} , \;\;  I = 2,3.$
  The total action is written as
\beqn
\label{eq:vvaym}
 S &=&
 \frac{1}{4 g^{2}} \;  Tr \left(
 \int d^2 \th W^{\a} W_{\a} + h.c. +
4 \int d^2 \th d^2 \thb \Phi^{\dg}_{I} e^{2V} \Phi_{I} e^{-2V} \right) \\
  &+&  \frac{1}{g^{2}} \sum_{f=1}^{n_f}
  \int d^2 \th d^2 \thb
\left( Q_{(f)}^{*} \left( e^{2V} \right) Q_{(f)}
+  \tilde{Q}_{(f)} \left( e^{-2V} \right) \tilde{Q}_{(f)}^{*} \right)
  + \frac{1}{g^{2}} \left( \int d^2\th W ( \th) + h.c. \;\right)\;,
\nonumber
\eeqn
 where the superpotential  is
\beqn
 W ( \th)  = \sqrt{2} Tr \left(
   \Phi_{1} \left[ \Phi_{2},  \Phi_{3} \right] \right) 
        +  \sum_{f=1}^{n_{f}} \left( m_{(f)} \tilde{Q}_{(f)} Q_{(f)}
        + \sqrt{2} \tilde{Q}_{(f)} \Phi_{1}Q_{(f)} \right) \;\;\;. 
\eeqn
 
  It is of interest to to express $S$ as
\beq
   S= S_{close} + \Delta S \;\;.
\eeq
  Fot this, we need to
 render $S$ to its component form.
 Solving the equation for the $D$ term, we obtain
\beqn
  D = \left[ \Phi^{\dagger}_{I}, \Phi_{I}\right] -  \hat{\rho}_{-}
 \sum_{f=1}^{n_{f}} \left( Q_{(f)}  Q_{(f)}^{*} -
 \tilde{Q}_{(f)}^{*}   \tilde{Q}_{(f)} \right)  \;\;,
\eeqn
where we have placed the $USp$ vectors $Q_{(f)},\; \tilde{Q}_{(f)}$ and their
  complex conjugates in the form of dyad.  The $F$ terms  are such that
\beqn 
 - \delta W = \sum_{I=1,2,3} tr F^{\dagger}_{\Phi_{I}} \delta \Phi_{I}
      + F^{*}_{Q_{(f)}} \delta  Q_{(f)} 
+ F^{*}_ {\tilde{Q}_{(f)}}  \delta \tilde{Q}_{(f)} \;\;\;.
\eeqn
  Explicitly
\beqn
\label{eq:F}
  F^{\dagger}_{\Phi_{1}} &=&  - \sqrt{2} \left[ \Phi_{2}, \Phi_{3}\right]
  - \sqrt{2} \hat{\rho}_{-} \left( \sum_{f=1}^{n_{f}} Q_{(f)}
 \tilde{Q}_{(f)} \right),\;
 F^{\dagger}_{\Phi_{2}} =  - \sqrt{2} \left[ \Phi_{3}, \Phi_{1}\right],
 \;\; 
 F^{\dagger}_{\Phi_{3}} = - \sqrt{2} \left[ \Phi_{1}, \Phi_{2}\right] \;,
  \nonumber \\
 F^{*}_{Q_{(f)}} &=& - \left( m_{(f)} \tilde{Q}_{(f)} + \sqrt{2}
 \tilde{Q}_{(f)} \Phi_{1} \right) \;,\;\;
  F^{*}_{\tilde{Q}_{(f)}} = - \left( m_{(f)} Q_{(f)} + \sqrt{2} \Phi_{1} 
 Q_{(f)} \right) \;\;.
\eeqn
 As for the Yukawa couplings, they can be read off from
\beqn
 \delta^{2} W \equiv  \sum_{A,B} \frac{\partial^{2}W}{\partial A \partial B}
\delta A \delta B \;\;,
\eeqn
  where the summmation indices $A,\;B$ are over all chiral superfields
 $\Phi_{I}\; I=1,2,3,$ and $Q_{(f)}, \tilde{Q}_{(f)}\;, \;
 f= 1, \cdots n_{f}$.

 Using the  complex $2n_{f}$ dimensional vectors and spinors introduced before,
  we find,  after some algebras,
\beqn 
   \Delta S &=& \Delta S_{b} + \Delta S_{f} =
 \left( S_{g-s} + {\cal V}_{scalar} + S_{mass} \right)
   + \left( S_{g-f} + S_{Yukawa} \right)  \;\;,  \\
 S_{g-s} &=&  -\frac{1}{g^{2}} tr \left( \sum_{\nu = 0,1,2,3,4,7}
  v_{\nu} v^{\nu} +  \sum_{I=2,3} \left[ \Phi_{I}, \Phi^{\dagger I} \right] 
  \right)   {\bf Q} \cdot {\bf Q}^{*}   \nonumber \\
     &+&  \frac{1}{g^{2}} tr \left[ \Phi_{2}, \Phi_{3} \right] F^{-1}
 {\bf Q}^{*} \cdot \Sigma  {\bf Q}^{*}
  - \frac{1}{g^{2}} tr \left[ \Phi_{2}^{\dagger}, \Phi_{3}^{\dagger}
 \right]
 {\bf Q} \cdot \Sigma F {\bf Q} \;\;,  \\
 S_{mass} &=& 
  - \frac{1}{g^{2}} tr \left( {\bf Q} \cdot M^{2} {\bf Q}^{*} \right)
  -\frac{2}{g^{2}} tr \left(v_{4} {\bf Q} \cdot M {\bf Q}^{*} \right)\;\;, \\
   {\cal V}_{scalar} &=&   - \frac{1}{2g^{2}} tr {\bf Q} \cdot
  \Sigma {\bf Q} 
  {\bf Q}^{*} \cdot  \Sigma  {\bf Q}^{*} 
 - \frac{1}{8g^{2}} tr \left[
 {\bf Q} \cdot {\bf Q}^{*} - F^{-1} {\bf Q}^{*} \cdot {\bf Q} F \right]^{2}
 \;,  \\
  S_{g-f} &=& \frac{1}{g^{2}} \left\{  \psivector^{\ast}
  \overline{\sigma}^{m} v_{m} \cdot \psivector
 + i \sqrt{2} {\bf Q}^{*} \lambda \cdot
  \psivector  - i \sqrt{2} \psivector^{\ast}
 \overline{\lambda} \cdot {\bf Q}   \right\} \;,  \\
 S_{Yukawa} &=& - \frac{1}{g^2}\left\{ \sum_{(c_{1}, c_{2})= 
   (Q, \tilde{Q}), (Q, \Phi_{1}), (\Phi_{1},\tilde{Q})}
 \frac{\partial^{2} W_{matter}}
{\partial C_{1} \partial C_{2}} \psi_{C_{2}} \psi_{C_{1}} + h.c. \right\}
   \nonumber \\
 &=& \frac{1}{g^2}\left( \frac{1}{2}   \psivector  \cdot  \Sigma   F
  \left( \sqrt{2} \Phi_{1} + M \right) \psivector
 + \sqrt{2} {\bf Q} \cdot  \Sigma  F \psi_{\Phi_{1}}  \psivector
  +  h.c. \right) \;\;.
\eeqn
  Here
\beqn
  \Sigma  &\equiv&  \left(
 \begin{array}{cc}
   0 & I  \\
   I & 0  
  \end{array}   \right)\;\;,  \\
M &\equiv&  {\rm diag} \left( m_{(1)}, \cdots, m_{(n_f)}
 - m_{(1)}, \cdots, -m_{(n_f)} \right) \;\;,  \\
  W_{{\rm matter}} &=&  \sum_{f=1}^{n_{f}} \left( m_{(f)}
 \tilde{Q}_{(f)} Q_{(f)}+ \sqrt{2} \tilde{Q}_{(f)} \Phi Q_{(f)} \right)\;\;,
\eeqn
  and $\cdot$ implies the standard inner product with respect to the
  $2n_{f}$ flavour indices.

\subsection{Notion of spacetime points}
 
    We have seen the validity  and the rationales of the 
  $USp(2k)$ reduced matrix model
 as a constructive  model descending from type $I$ superstrings.
 As we will see, the model is also dynamical theory of
 spacetime points.. 
   As is mentioned at the introduction, the dynamical degrees of freedom
  of spacetime points $X_{M}^{(i)}$ are embedded in the matrices.
  We write
\beqn
  v_{M} = X_{M}+ \tilde{v}_{M}   \;\;\;,
\eeqn
  where $\tilde{v}_{M}$ are off-diagonal matrices.
  One can imagine integrating out the bosonic off-diagonals
 and the fermions.  This will give dynamics
 to the spacetime points.  The object we study is the effective action for
 the spacetime points \cite{IM,CIK,AIKKT,HNT}:
\beq
   \ln Z\left[ X_{M}^{(i)}  \right] \;\;.
\eeq
   It is sometimes instructive to study the case in which the spacetime points 
  are assumed to interact weakly and are widely separated.
 In this case we can approximate the $USp$ matrices by
   individual $SU(2)$ blocks. We will find below that this approximation
  has a direct connection to string theory results obtained
 from the worldvolume gauge theories  in various dimensions.
 
\section{Matrix $T$ dual and representation as worldvolume gauge theories}
  Before going into the subject of this section, let us note that
  the classical solution of the model which gives  vanishing action 
  is broken  by $Z_{2}$ for six of the adjoint directions.
   In fact, the commutator of $\tilde{\delta} \equiv \delta^{(1)}
  + \delta^{(2)}$  with itself closes into translation only for
  four of the antisymmetric matrices as is clear from eq. (\ref{eq:am}).
 This in fact means the presence of
  the $O3$ fixed surfaces.
  The discussions in later sections  convince ourselves of
  the presence of $D3$ branes as well in the original representation.
 These appear in such a way  
 that the total $RR$ flux of the system cancels \cite{Pol}.

\subsection{general remarks and the series of degenerate vacua associated
 with  the model}
  What we would like to put forward with the reduced matrix model is
  a constructive approach  to unified string theory.  We presume  that
  the lifting of perturbative vacua requires genuinely nonperturbative
  mechanism and this can only be accomplished by  finding  the proper scaling limit
  of the model, which is a difficult task at this moment.

   As a separate theme of research,
   we try to establish  the consistency of the $USp$ matrix model with
   string  perturbation theory in the presence of D-branes
 (semiclassical object) via the worldvolume representation.
  In particular, one can think of toroidal compactifications in various
  dimensions  via the recipe of \cite{WT}.
   While some nonperturbative effects can be
 seen as exact results on worldvolume gauge theories, this representation
  leaves aside the original goal of lifting perturbative string vacua.
   In fact, one naively sets $k\rightarrow
   \infty $  and  has no hope of capturing the true vacuum.

 The procedure of matrix toroidal compactification  is well known
   and will not be reviewed here.  Instead, we just tabulate  the relevant
  properties  and the correspondence with  the ``classical'' counterparts.
  We have seen  in the last section that $F$ is a matrix counterpart of
 the twist operation $\Omega$.
  The matrix $T$ dual is nothing but a Fourier transform   $\hat{T}$.
  We can consider the combined transformation  of these. 
 We  have observed in \cite{IT2}
  that the sign flip occurs provided the worldvolume
  gauge fields are odd under parity.
  These are summarized as a table.

\noindent
 
\begin{tabular}{|c|c|c|}
\hline
 & "classical" & matrices \\
\hline
twist & $\Omega$ & $F$ \\
\hline
O3 fixed & $X_{\nu}(\bar{z}, z) = - \Omega X_{\nu}(z, \bar{z})
\Omega^{-1}$ &
$v_{\nu}^{t} = - Fv_{\nu}F^{-1}$ \\
surface & $X_{I}(\bar{z}, z) = + \Omega X_{I}(z, \bar{z}) \Omega^{-1}$ &
$v_{I}^{t} = + Fv_{I}F^{-1}$ \\
\hline
$T$ transf & $\hat{T}[X_{M}] \equiv X_{M \ \mbox{R}}(z) - X_{M
 \mbox{L}}(\bar{z})$ &
$\hat{T}((v_{M})_{\vec{a}, \vec{b}}) = <x|\hat{v}_{M}|x>$ \\
 & sign flip & sign flip OK \\
 & under $\Omega T$ & if $\tilde{v}_{\nu}(\vec{x}) = -
\tilde{v}_{\nu}(\vec{x})$ \\
\hline
\end{tabular}

\noindent

  From  matrix toroidal compactification,
  we find a series of degenerate perturbative vacua associated with
  the $USp(2k)$  matrix model.
In the remainder of this section, we will see the consistency
 of the worldvolume representation of the $USp$ matrix model in various
 dimensions  with some  literature.  When some of the adjoint
  directions get compactified and become small, it is preferrable to  T-dualize
  the system into these diretions.
  One can imagine this as a figure:

\begin{figure}[t]
\centerline{\epsfbox{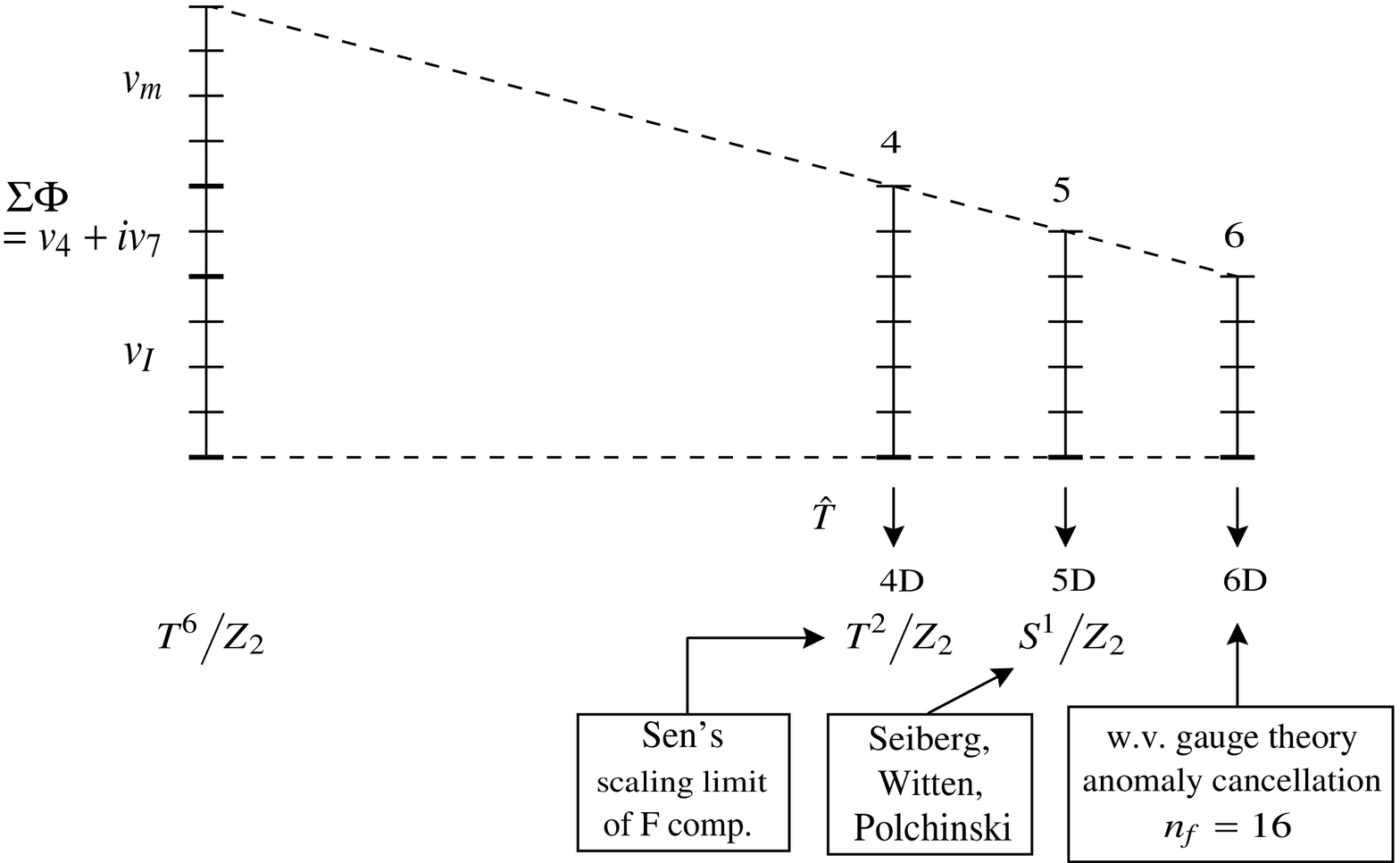}}
\label{fig3}
\end{figure}
 
\subsection{6d,5d,4d worldvolume representations}
    We now look at the $T$ dualized form of the $USp$ matrix model
  for the cases of the $6d,5d,4d$ worldvolume theories, which
  respectively represent  type$I$ theory 
  in  ten dimensions, its orientifold compactification
  on  $S^{1}/Z^{2}$  ($9d$ string theory)  and that on $T^{2}/Z^{2}$
  ($8d$ string theory).  We will demand that all of the
degenerate perturbative vacua discussed above be infrared stable.
 We  will find out
$n_{f}=16$ and  in each case we distribute  these evenly around
the fixed surfaces.

\subsubsection{6d}
 
 For consistency, we have added the fields lying in
 the fundamental representation.
 They are responsible for creating an open string sector.
 In string perturbation theory,   the infrared stability is seen
  through the (global) cancellation of dilaton tadpoles between disk and
 $RP^{2}$ diagrams \cite{GS}, \cite{IMo}, leading to the $SO(32)$ Chan-Paton
 factor. This survives toroidal compactifications with/without discrete
 projection \cite{AIKT}. 

  We have found out that in the case of all six adjoint directions
 compactified,  the infrared stability gets translated into
  the consistency of the  six dimensional worldvolume  $USp(2k)$ gauge theory
 with matter in the antisymmetric and fundamental representations. 
   In fact, by acting 
\beqn
\Gamma_{(6)} \equiv \Gamma^{0}\Gamma^{1}\Gamma^{2}\Gamma^{3}
\Gamma^{4} \Gamma^{7} 
\eeqn 
 on $\Psi$, we see that the adjoint fermions $\lambda$ and $\psi_{1}$ have
 chirality plus  while $\psi_{2,3}$ have chirality minus.  The fermions in the
 fundamental representation have chirality minus.
 The standard technology
to compute nonabelian anomalies is provided by family's index theorem
 and the descent equations.
 We find that the condition  for the anomaly cancellation:
\beqn
   & &  tr_{adj}F^{4} - tr_{asym} F^{4} - n_{f} tr F^{4}  \;\; \nonumber \\
& & = (2k+8) trF^{4} + 3 \left( trF^{2}\right)^{2}  
  - \left( \left(2k-8\right)  trF^{4} + 3 \left( trF^{2}\right)^{2}
  \right)
  - n_{f} trF^{4}   \nonumber \\
& & = \left( 16-n_{f} \right) tr F^{4} =0 \;\;\;,
\eeqn
  where we have indicated the traces in the respective representations.
 The case $n_{f}=16$ is selected by the consistency of the theory.
In the case discussed in eq. (\ref{eq:2-8 case projector}),
we conclude from similar calculation that
the anomaly cancellation of the worldvolume two-dimensional
gauge theory selects sixteen complex fermions.

 This result is  reasonable  from the ``classical'' consideration
  of the $RR$ flux.  We know that the flux is from  $O3$  to $n_{f}$  of
  $D3$ and their mirror  and is in the adjoint directions.
  When all  of the six adjoint directions  are compactified,
  the flux cannot escape to infinity  and  the total flux had better vanish.
  In this way, we also get $n_{f}=16$. 
Although we will not discuss here, a version of anomaly inflow argument is 
  operating  which relates the conservation of the $RR$ flux to
   the nonabelian anomaly.

\subsubsection{space-dependent (axion-) dilaton background field}

 The simplest quantity to be computed in the worldvolume representation of
 matrix models in general is the effective running coupling constant
 $g^{2}_{eff}(u)$ obtained from the low-energy effective action.
 Here $\vec{u}$ is a $vev$ of the scalars which labels quantum moduli space. 
 The $g^{2}_{eff}(\vec{u})$ represents the marginal scalar
 deformation of the original action  to a type of nonlinear $\sigma$ model.
  The background field appearing through this procedure is a massless
 (axion-)dilaton field. The running coupling constant is
 identified as the space-dependent (axion-) dilaton background field.
  One strategy to compute  $g^{2}_{eff}(u)$ is
 $i)$ to begin with computing  the part of the effective action associated with 
  the fermionic integrations, $ii)$ to invoke supersymmetry to find
out $S_{eff}$, and $iii)$ to take the hessian of $S_{eff}$ to obtain
 $g^{2}_{eff}(u)$.

\subsubsection{5d}
 
 Let us first consider the widely separated case, namely,
  $SU(2)$ case  with eight flavours per fixed surface.  
$X_{7}=u$ and $X_{5,6,8,9}$ decouple.
  Omitting the  derivation, we  write the result \cite{Seiberg}
\beqn
  \frac{1}{g^{2}_{eff} (u)} =  \frac{1}{g^{2}} + 16X_{7}
- \sum_{f=1}^{8} \mid X_{7}- m_{f} \mid -
   \sum_{f=1}^{8} \mid X_{7}+ m_{f} \mid  \;\;.
\eeqn
(See also \cite{PW}.)
 The $USp(2k)$ matrix model provides a natural generalization
  to this.   The answer  this time depends on the $k$ spacetime points
 $X_{7}^{(i)}$   and on those of the antisymmetric directions  
 $X_{5,6,8,9}^{(i)}$.  We  have completed  the first step of the
 computation \cite{AI} and the answer appears to reflect that the
 spacetime is curved in the  $X_{5,6,8,9}^{(i)}$ directions.
 
\subsubsection{4d}
It has been shown  that the model is able to describe
\cite{IT1,IT2} the $F$ theory compactification on an elliptic fibered
$K3$ \cite{V,Sen}.   We will not repeat the discussion here.
In the widely separated case
($SU(2)$ with four flavours and 
$u= X_{4}+ iX_{7}$ ), the $4d$ worldvolume theory  obtaind
is exactly Sen's scaling limit to $F$ on $K3$ \cite{Sen}. 
The space-dependent
axion/dilaton background field is controlled by the Seiberg-Witten curve.
The model again provides an interesting generalization to
the $USp(2k)$ case and much remains to be worked out.

\section{Formation of extended objects from (non)-abelian Berry phase}
     We now show that the model contains  degrees of freedom
 corresponding to D-objects. We study the effective action for the
 spacetime points.
  In particular, we study the effects of fermionic integrations
  which contain the information of the $RR$ sector  to deduce the
 coupling of D-objects  to spacetime.   This will be the effect which
 survives the cancellation of the bosonic
  integrations against the fermionic ones.  We will suppress here the
 nature of time in the reduced model as  dynamical variable in order to
 form  a loop in the parameter space ($=$ spacetime points.)
 We will study the coupling in the representation of the model as
 $T$ dualized quantum mechanics via the Berry phase.
 
\subsection{path-dependent effective action}
    We will make the effective action dependent on
 the paths $\{ \Gamma_{A}^{(R)} \}$ in the parameter space
  labelled by the five sets of the adjoint spacetime points
$X_{\nu} = diag(X_{\nu}^{(1)}, \cdots X_{\nu}^{(k)},$
$ -X_{\nu}^{(1)}, \cdots -X_{\nu}^{(k)}.)$ $\nu =1,2,3,4,7$.
We choose $v_{0}=0$ gauge.
\beqn
\label{eq:adprocess}
Z \left[ X_{\nu}; m_{(f)}, \{\Gamma_{A}^{(R)} \},
  \{ \sigma_{i A}^{(R)} \},  \{ \sigma_{f A}^{(R)} \}
 \right] =  \int \left[ D \tilde{v}_{M} \right]    \prod_{f=1}^{n_{f}}
\left[ D Q_{(f)}\right] \left[ D Q_{(f)}^{*}\right]   
  \left[D {\tilde{Q}}_{(f)} \right] \nonumber \\
\left[ D \tilde{Q}_{(f)}^{*} \right]  e^{iS_{B}} 
 \lim_{T \rightarrow \infty} \langle  \Psi ; \{ \sigma_{i A}^{(R)} \}  
  \mid P \exp
 \left[ -i  \int_{0}^{T} dt H_{fermion}(t) \right]
 \mid \Psi ; \{ \sigma_{f A}^{(R)} \}  \rangle \;. 
\eeqn
  Here $v_{M} = X_{M}+ \tilde{v}_{M}$ and $S^{B}$ is the pat of $S$
which does not contain any fermion. For simplicity, we have set the
 dependence of the remaining antisymmetric spacetime points
 $X_{I}=  diag(X_{I}^{(1)},
 \cdots X_{I}^{(k)},$$ X_{I}^{(1)}, \cdots X_{I}^{(k)})$ $\;I= 5,6,8,9$ 
  to zero.
 The operator $H_{fermion}$  is the sum of the respective
 Hamiltonians $H_{{\rm fund}},H_{{\rm adj}}$ and $H_{{\rm asym}}$ obtained
 from the fermionic part of
$ S_{{\rm fund}}, S_{{\rm adj}}$ and $S_{{\rm asym}}$ after $T$ duality.
Their $t$ dependence comes from that of $X_{\nu}$  which acts as external
parameters on the Hilbert space of fermions. 
We consider a set of degenerate
eigenstates. The degeneracy of the initial state and that of the final one
are respectively specified by  a set of labels  $\{ \sigma_{i A}^{(R)} \}$
and $\{ \sigma_{f A}^{(R)} \}$, where the indices $A$ and $(R)$  specify
 the species of fermions.
 
\subsection{reduction to the first quantized problem}  
  Let $R = {\rm fund, adj, antisym}$. We denote by $e_{(R)}^{(A)}$
the standard eigenbases belonging to the roots of $sp(2k)$ and the weights
 of the fundamental representation and those of the antisymmetric
representation respectively.
Let us expand the two component fermions as
\beqn
\label{eq:expand}
 \psi^{(R)} = \sum_{A}^{N_{(R)}} b_{A}^{(R)}
 e_{(R)}^{(A)} /\sqrt{2} \;,\;\;
\bar{\psi}^{(R)} = \sum_{A}^{N_{(R)}} \bar{b}_{A}^{(R)}
 e_{(R)}^{(A) \dagger}/ \sqrt{2} \;\;\;,
\eeqn
  where  $N_{({\rm adj})}= 2k^{2} +k$,
  $N_{({\rm antisym})}= 2k^{2} -k$ and  $N_{({\rm fund})}= 2k$.
  We find that all of the three Hamiltonians
 $H_{{\rm fund}},H_{{\rm adj}}$ and $H_{{\rm asym}}$
 are expressible in terms of  a generic one
\beqn
\label{eq:abelian}
  g^{2} H_{0} \left(X_{\ell}, \Phi, \Phi^{*}; (R), A \right) =
-\bar{b}_{A\dot{\a}}^{(R)} {\s}^{m\dot{\a}\a}X_{m}b_{A\a}^{(R)}
-d^{(R)\a}_{A} \s_{\a\dot{\a}}^{m}X_{m} \bar{d}_{A}^{(R) \dot{\a}}
+ \sqrt{2}  \Phi b^{(R) \a}_{A}  d_{A\a}^{(R)}  \nonumber \\
+\sqrt{2} \Phi^{*}\bar{b}_{A\dot{\a}}^{(R)}
  \bar{d}^{\dot{\a} (R) }_{A} \;\;
\eeqn
provided we replace the five parameters
\beq
X_{\ell},\;\;\; \Phi = \frac{X_4 + iX_7}{\sqrt{2}}, \;\;\;
\Phi^{*} = \frac{X_4 - iX_7}{\sqrt{2}} \;\;\; \ell =1,2,3
\eeq
by the appropriate ones. (See argument of ${\cal A}$ in
eq.~(\ref{eq:answer1}) below.)

The Berry connection appears in one or three particle
 state of $H_{0}$ with respect to
 the Clifford vacuum $\ket{\Omega}$;
 $b^{\a}\ket{\Omega} = \bar{d}_{\dot{\a}}\ket{\Omega} = 0$.
  We suppress the labels $A$ and $(R)$ seen in
 eqs. (\ref{eq:expand}),(\ref{eq:abelian}) for a while.
  Let us write   $\mid \Psi  \rangle =  \left( h_{\a} d^{\a}
+ \bar{h}^{\dot{\a}} \bar{b}_{\dot{\a}}  \right) \mid \Omega \rangle $ and
 $\psi \equiv \left( h_{\a} , \bar{h}^{\dot{\a}} \right)^{t}$.
 The transition amplitude reads
\beqn
\label{eq:formula}
 \lim_{T \rightarrow \infty} \langle \Psi \mid P \exp \left[ -i  
\int_{0}^{T}
 dt H_{0}(t) \right] \mid  \Psi \rangle
  =  \psi^{\dagger}
 P \exp \left[ -i \int_{0}^{\infty} E(t) dt
 + i \int_{\Gamma} d\gamma (X_{m}, \Phi, \Phi^{*})  \right]
 \psi\;.
\eeqn
Here $\Gamma$ is a path in the parameter space. The  
connection one-form  is
\beqn
  id\gamma (t)= -  \psi^{\dagger} (t) d \psi (t)
  \equiv   -i {\cal A} \;\;\;,
\eeqn
  which is in general matrix-valued.
 Let us consider for definiteness a set of two degenerate adiabatic
 eigenstates with positive energy, which is specified by an index
 $\sigma =1,4$.  Using the completeness
  $ {\displaystyle \sum_{\sigma =1,4} } \psi_{\sigma}
 \psi_{\sigma}^{\dagger} = {\bf 1}_{(2)}$,
   we find
\beq
\label{eq:factor}
\sum_{\sigma = 1,4}  \lim_{T \rightarrow \infty} \langle \Psi_{\sigma}
 \mid P \exp \left[ -i \int_{0}^{T}
 dt H_{0}(t) \right] \mid  \Psi_{\sigma} \rangle =
  tr  P \exp \left[ -i \int_{0}^{\infty} E(t) dt
 - i \int_{\Gamma} {\cal A}(X_{\nu}) \right] 
\eeq
 from eq. (\ref{eq:formula}).
  Here the trace is taken with respect to the two-dimensional subspace.

\subsection{computation of the Berry phase and the BPST instanton}
  Now the problem is to obtain the nonabelian $(su(2)$ Lie algebra valued)
  Berry connection associated with the fist quantized hamiltonian
\begin{equation}
{\cal H} = \frac{R}{g^2} \sum_{\nu=1,2,3,4,7}  N^{\nu}  \Gamma_{\nu}\;, \;
   R \equiv \sqrt{(X^1)^2 +(X^2)^2 +(X^3)^2 +(X^4)^2 +(X^7)^2}\;,
 \; N^{\nu} \equiv \frac{X^{\nu}}{R} \;,
\end{equation}
  where $\Gamma_{\nu}$ are the five dimensional gamma matrices obeying
 the Clifford algebra and the explicit representation can be read off
  from eq.~(\ref{eq:abelian}).
  The projection operators are
\begin{equation}
P_{\pm} = \frac{1}{2} ({\bf 1}_4 \pm  N^{\nu} \Gamma_{\nu}) \;\;, \;\;
{P_{\pm}}^2 = P_{\pm} \;\;,\;\; 
 P_+^{\dag} = P_+\;\;, 
\end{equation}
   which satisfy
\begin{equation}
{\cal H} P_{\pm} = \pm \frac{R}{g^2} P_{\pm}\;\;.
\end{equation}
  Denoting by ${\bf e}_i$ $(i=1,2,3,4)$ the unit vector in the $i$-th
 direction,  we write a set of normalized eigenvectors belonging to
 the plus eigenvalue as
\begin{equation}
 \psi_i  = \frac{1}{{\cal N}_i} P_{+}{\bf e}_i \;\;.
\end{equation}
 Here, $i=1,4$ refer to the sections around the north pole  $X^{3}=R$ 
  while  $i=2,3$  to the ones around the south pole $X^{3}=-R$.
  The ${\cal N}_i$ are the normalization factors:
\begin{equation}
{\cal N} \equiv {\cal N}_1 = {\cal N}_4 = \sqrt{\frac{1+N^3}{2}}\;\;,\;\;
 {\cal N}^{\prime} \equiv{\cal N}_2 = {\cal N}_3 = \sqrt{\frac{1-N^3}{2}}\;\;.
\end{equation}
  We focus our attention on the sections near the north pole. The Berry
 connection  is
\begin{equation}
i{\cal A}  = \left( \begin{array}{l}
			  \psi_1, \\
			 \psi_4, 
		\end{array} \right)
	 d (  \psi_1,  \:  \psi_4  )
\end{equation}

  We will here present the final answer only.
We parametrize  $S^{3}$ of unit radius by the coordinates
\begin{equation}
Y^{\nu} \equiv \frac{1}{\sqrt{R^2 - (X^3)^2}} X^{\nu}\:\:,\:\:\:
 (\nu = 1,2,4,7)
\end{equation}
\begin{equation}
(Y^1)^2 + (Y^2)^2 + (Y^4)^2 + (Y^7)^2 = 1\;\;.
\end{equation}
   Also let ${\bf Y} \equiv ( Y^4, Y^7, Y^1 )^{t}$.
  The $Y$ coordinates parametrize the $SU(2)$ group element as well:
\begin{equation}
T \equiv Y^2 {\bf 1}_2 + i {\bf Y} \cdot \mbox{\boldmath$\sigma$}\;\;,
\end{equation}
  from which  we can make the pure gauge configuration
\begin{equation}
\begin{array}{rcl}
dT T^{-1} 
 &=& (dY^2 {\bf 1}_2 + i d{\bf Y} \cdot \mbox{\boldmath$\sigma$})
(Y^2 {\bf 1}_2 - i {\bf Y} \cdot \mbox{\boldmath$\sigma$})\\
 &=& i (d{\bf Y} \times {\bf Y} + Y^2 d{\bf Y} -{\bf Y} dY^2) \cdot 
\mbox{\boldmath$\sigma$}.
\end{array}
\end{equation}
  The final answer we have found out is 
\begin{equation}
\label{eq:BPST}
{\cal A} \left( X^{\nu} \right) = p(R, X^{3}) dT T^{-1}\;\;.
\end{equation}
 The prefactor $p(R, X^{3})$ is of interest and can be written as
\beq
 p(R, X^{3}) =  \frac{\tau^2}{\tau^2 + \lambda^2}  \;\; , \;\;\;
\tau = \sqrt{R^2 - (X^3)^2}\;\;,\;\;\;
\lambda = R + X^3.
\eeq
 The nonabelian connection ${\cal A}$ is in fact the BPST instanton
 configuration.  The size of the instanton
 $\lambda$
 is not a ${\it bonafide}$ parameter of the model but is chosen to be
 the fifth
 coordinate in the five dimensional Euclidean space.  For fixed $\lambda$, 
 the four dimensional subspace embedded into  the ${\bf R}^{5}$  is a 
 paraboloid wrapping the singularity. An observer on this recognizes
 the pointlike singularity as the BPST instanton.  As $\lambda$ goes to zero,
 this paraboloid gets degenerated into an $SU(2)$ counterpart of the Dirac
 string connecting  the origin and the infinity.
     Note also that the prefactor is written in terms of the angle measured
 from the north pole as
\begin{equation}
p(\theta) = \frac{1}{2}(1- \cos \theta )\;\;,\;\;
N^3  \equiv R \cos \theta \;\;.
\end{equation}
 
\subsection{coupling to spacetime points}
 Returning to the expression (\ref{eq:adprocess}) and
 taking a sum over the labels $\sigma_{f A}^{(R)} = \sigma_{i A}^{(R)}$,
  we find that the second line is expressible as the product of the factors
\beqn
\label{eq:answer1}
 &&  Tr P \exp \left( -i \sum_{f=1}^{n_{f}} \sum_{A=1}^{2k}
 {\bf 1} \otimes \cdots \int_{\Gamma_{A,f}^{(fund)} } {\cal A}
\left[ {\bf w}^{A}\cdot {\bf X} _{\ell},\;
 \frac{ m_{(f)}}{\sqrt{2}}    +{\bf w}^{A} \cdot {\bf\Phi},\;
 \frac{ m_{(f)}}{\sqrt{2}} + {\bf w}^{A} \cdot {\bf \Phi }^{\dagger}
 \right] \cdots \otimes {\bf 1} \right) 
   \nonumber \\
  &&  Tr  P exp \left( -i \sum_{A=1}^{2k^{2}}
   {\bf 1} \otimes \cdots \int_{\Gamma_{A}^{(adj)} } 
  {\cal A} \left[ {\bf R}^{A} \cdot {\bf X}_{\ell},\;
 i {\bf R}^{A} \cdot {\bf \Phi},\;
  i {\bf R}^{A} \cdot {\bf \Phi}^{\dagger} \right]  \cdots
 \otimes {\bf 1} \right)  \nonumber \\  
 &&  Tr  P exp \left( 
 -i  \sum_{A=1}^{2k^2 - 2k} {\bf 1} \otimes \cdots 
 \int_{\Gamma_{A}^{(asym)} }
 {\cal A} \left[ {\bf w}_{{\rm asym}}^{A} \cdot {\bf X}_{\ell},\;
 {\bf w}_{{\rm asym}}^{A} \cdot {\bf \Phi},\;
  {\bf w}_{{\rm asym}}^{A} \cdot {\bf \Phi}^{\dagger} \right] \cdots
 \otimes {\bf 1} \right) \;. 
\eeqn
   We have included  the energy dependence seen in eq. (\ref{eq:factor})
 in $S_{B}$ as  this is perturbatively cancelled by the contribution from
 bosonic integration.  The symbols seen in the arguments are
\beqn
\{ \{  {\bf w}^{A} \mid  1 \leq A \leq 2k  \}\}
&=&
\{ \{   \pm {\bf e}^{(i)}
\; , 1 \leq i \leq k   \}\}   \;\; \nonumber \\
\{ \{ {\bf R}^{A} \mid 1 \leq A \leq 2k^{2} \}\}
&=&
\{ \{   \pm 2 {\bf e}^{(i)}, {\bf e}^{(i)}-{\bf e}^{(j)} ,
\pm \left({\bf e}^{(i)} +{\bf  e}^{(j)} \right)
\; 1 \leq i,j, \leq k   \}\}  \;\; \nonumber \\
\{ \{  {\bf w}_{ {\rm asym} }^{A} \mid  1 \leq A \leq 2k^{2} -2k   
\}\}
&=&
\{ \{    \pm \left( {\bf e}^{(i)} +{\bf  e}^{(j)}  \right) ,
{\bf e}^{(i)} - {\bf e}^{(j)},  \; 1 \leq i,j, \leq k   \}\} \;\;.
\eeqn	
  The second and the third lines
 are respectively the nonzero roots and the weights in the  
antisymmetric representation
 of $usp(2k)$. We have denoted by ${\bf e}^{(i)} \;(1 \leq i \leq  k)$
the orthonormal basis vectors of $k$-dimensional Euclidean space and
\beqn
{\bf X}_{\ell} = \sum_{i=1}^{k}  {\bf e}^{(i)} X_{\ell}^{(i)},\;
{\bf \Phi} =  \sum_{i=1}^{k}{\bf e}^{(i)}
\frac{X_{4}^{(i)} + iX_{7}^{(i)}}{\sqrt{2}}, \;
{\bf \Phi}^{\dagger}= \sum_{i=1}^{k}{\bf e}^{(i)}
\frac{X_{4}^{(i)} - iX_{7}^{(i)}}{\sqrt{2}}
\;\;\;.
\eeqn

 Let us exploit the symmetry of the roots and the weights
under ${\bf e}^{(i)} \leftrightarrow -{\bf e}^{(i)}$.
  In general, neither the second line nor the third one collapse to unity. 
 Observe that, due to this symmetry, we can pair
 the two-dimensional vector space associated with $ {\bf R}^{A}$
  (or ${\bf w}_{ {\rm asym} }^{A}$) and that with   $-{\bf R}^{A}$
 (or $-{\bf w}_{ {\rm asym} }^{A}$).  Let us symmetrize the tensor product
 of these two two-dimensional vector spaces. On this, the nonabelian
 Berry phase is reduced to the pure gauge configuration
\beq
 {\cal A}(X^{\nu})_{ \{i }^{ \;\;\{j } \delta_{k \} }^{\;\; \ell \} }
 + {\cal A}(-X^{\nu})_{ \{i }^{ \;\;\{j } \delta_{k \} }^{\;\; \ell \} }    
 = \left( dT T^{-1} \right)_{ \{i }^{ \;\;\{j }
 \delta_{k \} }^{\;\; \ell \} }   \;\;\;,
\eeq   
 and this can be gauged away.
  As for the first line of eq. (\ref{eq:answer1}), the mass terms prevent
 this from happening.

  After all these operations, eq. (\ref{eq:answer1}) becomes
\beqn
\label{eq:answer2}
   Tr P \exp \left( -i \sum_{f=1}^{n_{f}} \sum_{A=1}^{2k}
 {\bf 1} \otimes \cdots \int_{\Gamma_{A,f}^{(fund)} } {\cal A}
 \left[ {\bf w}^{A}\cdot {\bf X} _{\ell},\;
 \frac{m_{(f)}}{\sqrt{2}}    +{\bf w}^{A} \cdot {\bf\Phi},\;
 \frac{m_{(f)}}{\sqrt{2}} + {\bf w}^{A} \cdot {\bf \Phi }^{\dagger}
 \right]  \cdots \otimes {\bf 1} \right) \; \nonumber \\
     \;\;
\eeqn
  Let us note that before the symmetrization, the nonabelian Berry phase
 is present in the $IIB$  case.

\subsection{abelian approximation}

  In order to understand better the formation of the BPST instanton
 obtained from the nonabelian Berry phase, we will study
  this problem ignoring the degeneracy  {\it i.e.} in the abelian
 approximation. The section then  takes  the tensor product form of two
 two-component wave functions 
\beqn
  \psi_{i} \equiv  \psi_{A, a}  \equiv
\left(
\begin{array}{c}
 h_{\alpha} \\
 \bar{h}^{\dot{\alpha}}
\end{array}
\right) \;\;.
\eeqn
  Separation of variables is done by  the following five
 dimensional spherical coordinates
\beqn 
X_2 &=& r \sin \phi_1 \sin \theta_1 \cos \theta_2  \;\;\;, \nonumber  
\\
X_1 &=& r \cos \phi_1 \sin \theta_1 \cos \theta_2 \;\;\;,  \nonumber  
\\
X_3 &=& r  \;\;\;\;\;\;\;\;\;     \cos \theta_1 \cos \theta_2
 \;\;\;, \nonumber \\
X_4  &=& r \;\;\;\;\;\;\;\;\;\;\;\;\;\;\;\;\;\;
 \sin \theta_2 \cos \phi_2  \;\;\;, \;\; \qquad 0 \leq \phi_2 \leq 2  
 \pi \;\;, \nonumber \\
X_7 &=& r   \;\;\;\;\;\;\;\;\;\;\;\;\;\;\;\;\;\;
 \sin \theta_2 \sin \phi_2  \;\;\;, \;\; \qquad 0 \leq \theta_2 \leq  
\pi \;\;.
\eeqn
 The local form of the section
 around $\left( \theta_{2},\theta_{1} \right)$ denoted by $(N,N)$
 is
\beqn
\label{eq:efn}
\psi_{A, a}^{(N,N)} =
\left(
\begin{array}{c}
 \sin \frac{\theta_2}{2} \, e^{-i \phi_2} \\
 \cos \frac{\theta_2}{2} \\
\end{array}
\right)_{A}
\otimes
\left(
\begin{array}{c}
 \cos \frac{\theta_1}{2} \\
 \sin \frac{\theta_1}{2} \, e^{i \phi_1} \\
\end{array}
\right)_{a}  \;\;\;.
\eeqn
 The connection one-form is
\beqn
\label{eq:BC}
{\cal{A}}^{(N,N)} =
- \frac{i}{2} ( 1 - \cos \theta_2 ) d \phi_2
 + \frac{i}{2} ( 1 - \cos \theta_1 ) d \phi_1   \;\;.
\eeqn
  We see that the BPST instanton is made of a monopole
 anti-monopole pair and the flux of the monopole and
  that of the antimonopole spread in the different
 directions of spacetime.

\subsection{brane interpretation}

  This time, cancellation noted before occurs without symmetrization.
The cancellation occurs as well to the part from the fundamental  
representation which does not involve $X_{4}^{(i)}$ or $X_{7}^{(i)}$.
We find that the exponent of eq.~(\ref{eq:answer2}) is
written as
\beqn
 && -i \sum_{f=1}^{n_{f}} \sum_{i=1}^{k} 
 \gamma_{\Gamma}^{({\rm Berry})}\left[  X_{3}^{\prime (i)},
  m_{f}+ X_{4}^{(i)},  X_{7}^{(i)} \right]
    \nonumber \\
 && -i \sum_{f=1}^{n_{f}} \sum_{i=1}^{k}
 \gamma_{\Gamma}^{({\rm Berry})}\left[  X_{3}^{\prime (i)},
m_{f}- X_{4}^{(i)},  X_{7}^{(i)}   \right] \;\;.
\eeqn
Here
\beqn
\gamma_{\Gamma}^{({\rm Berry})} \left[ X_{3}^{\prime}, X_{4}, X_{7}
\right]
  &=&  \int {\cal A}^{({\rm Berry})} \\  
{\cal{A}}^{(N)({\rm Berry})} &=&
- \frac{i}{2} ( 1 - \cos \theta_2 ) d \phi_2 \;, \;\;
{\cal{A}}^{(S)({\rm Berry})} =
+\frac{i}{2} ( 1 + \cos \theta_2 ) d \phi_2 . \label{eq:config}
\eeqn
and $X_{3}^{\prime (i) 2} = X_{1}^{(i)2} + X_{2}^{(i)2}+ X_{3}^{(i)2}$.

    It is satisfying to see a pair of magnetic monopoles sitting at
$X_{4}^{(i)} = \pm m_{(f)}$ from the orientifold surface
for $i= 1 \sim k$.  These monopoles live in the
parameter space,
which is the spacetime points of the matrix model.
Coming back to eq.~(\ref{eq:adprocess}), we conclude that the Berry
phase generates an interaction
\beqn
\label{eq:induced}
Z \left[ x_{\ell}, x_{I}=0 ; \cdots\right] =
  \int \left[ D \tilde{v}_{M} \right]    \prod_{f=1}^{n_{f}}
\left[ D Q_{(f)}\right] \left[ D Q_{(f)}^{*}\right]
  \left[D {\tilde{Q}}_{(f)} \right]
\left[ D \tilde{Q}_{(f)}^{*} \right]  \exp [ iS_{B} +  
 i \gamma_{\Gamma}^{({\rm total})} ] . \nonumber
\eeqn

Let us give this configuration we have obtained a brane
interpretation  first from the six dimensional and
subsequently from the ten dimensional
point of view. It should be noted that the two coordinates
which the connection ${\cal{A}}^{({\rm Berry})}$ does not depend
on are the angular cooordinates $\theta_{1}, \phi_{1}$, so that 
$X_{1}, X_{2}$ are not quite separable from the rest of the
coordinates $X_{3}, X_{4}, X_{7}$  in
eq.~(\ref{eq:config}).  Only in the asymptotic region 
$\mid X_{3}^{\prime} \mid >> \mid X_{3} \mid$, there exists
an area of size $\pi \mid X_{3}^{\prime} \mid^{2}$ transverse to
the three dimensional space  where the Berry phase is obtained.
In this region, the magnetic flux obtained from the $b(=1)$-form
connection embedded in $d(=6)$-dimensional spacetime looks
approximately as is discussed in \cite{Nepomechie}: the flux
no longer looks coming from a poinlike object but from
a $d-b-3(=2)$ dimensionally extended object.
The magnetic monopole obeying the Dirac quantization behaves
approximately like a $D2$ brane extending to the
(1,2) directions, which are perpendicular to the orientifold surface.
In fact, the presence of this object and its quantized magnetic flux
have been detected by quantum mechanics of a point particle (electric
$D0$ brane) obtained from the $n=1$ and $n=3$ particle states of
the fermionic sector in the fundamental  representation.
The induced interaction is a minimal one. We conclude that the $D0$  
represented by the first and the third excited states of the quantum  
mechanical problem given above is under the magnetic field created
by $D2$. To include the four remaining coordinates
$(X_5, X_6, X_8, X_9)$ of the  antisymmetric directions, we appeal
to the translational invariance which is preserved in these
directions.  The simplest possibility is that they appear in the
coupling through the derivatives
\beqn
\int {\cal A}^{({\rm Berry})} =
\int\prod_{I=5,6,8,9} dX^{I}dX^{\nu} A_{\nu 5689}  \;\;.
\eeqn 
With this assumption,
the $D0$ brane is actually a $D4$ bane extended in $(5,6,8,9)$
directions while the  $D2$ still occupies $(1,2)$: the
quantization condition is preserved in ten dimensions as well.

\section{Schwinger-Dyson equations}
  We will here repeat the basic part of the discussion in \cite{Itsu}.
\subsection{derivation of S-D/loop equations}
 Let us derive S-D/loop equations, employing the open and the
 closed loop variables
 introduced in section $2.3$.
We first introduce abbreviated notation:
\beqn
\Phi[(i)] &\equiv& \Phi[p^{(i)}_{.},\eta^{(i)}_{.};n^{(i)}_{1},n^{(i)}_{0}]\;,
 \;\Psi[(i)] \equiv
\Psi_{f^{(i)'}f^{(i)}}[k^{(i)}_{.},\zeta_{.}^{(i)};l^{(i)}_{1},l^{(i)}_{0};
\Lambda^{(i)'},\Lambda^{(i)}]\;,  \nonumber\\
\int d\mu  \cdots &\equiv& \int [dv][d\Psi][d\Qvector][d\Qvector^{\ast}]
                        [d\psivector][d\psivector^{\ast}]  \cdots \;.
\nonumber 
\eeqn
We begin with the following set of equations consisting of $N$ closed loops 
and $L$ open loops:
\beqn
&&0=\int d\mu
\frac{\partial}{\partial X^{r}}
\left\{ Tr(U[p^{(1)}_{.},\eta^{(1)}_{.};n^{(1)}_{2},n^{(1)}_{1}+1] T^r 
U[p^{(1)}_{.},\eta^{(1)}_{.};n^{(1)}_{1},n^{(1)}_{0}]) \right. \nonumber\\
&& \hspace*{4.5cm}
\left.  \Phi[(2)] \cdots \Phi[(N)]  \Psi[(1)] \cdots \Psi[(L)] 
\, e^{-S} \right\} \;\;,
\label{loopeq1}
\\
&& 0=\int d\mu \frac{\partial}{\partial X^{r}}
\left\{  \left( \Lambda^{(1)'}  \Pivector_{(f^{(1)'})}  \right) 
F U[k^{(1)}_{.},\zeta^{(1)}_{.};l^{(1)}_{2},l^{(1)}_{1}+1]T^r
U[k^{(1)}_{.},\zeta^{(1)}_{.};l^{(1)}_{1},l^{(1)}_{0}]
 \left( \Lambda^{(1)} \Pivector_{(f^{(1)})}  \right) \right.  \nonumber\\
&&\hspace*{4.5cm}
 \left. \Phi[(1)] \cdots \Phi[(N)] 
\Psi[(2)] \cdots \Psi[(L)]
\, e^{-S} \right\} \;\;,
\label{loopeq2}
\\
&& 0=\int d\mu
\frac{\partial}{\partial \Zvector_{(f) i}}
\left\{ (U[k^{(1)}_{.},\zeta^{(1)}_{.};l^{(1)}_{1},l^{(1)}_{0}]
\left( \Lambda^{(1)}  \Pivector_{(f^{(1)})} \right)  )_{i}  \right. \nonumber\\
&& \hspace*{5cm}  \left. \Phi[(1)] \cdots \Phi[(N)]  \,
\Psi[(2)] \cdots \Psi[(L)]
\, e^{-S} \right\} \;\;,
\label{loopeq3}
\eeqn
where $X^r$  denotes $v_{M}^{r}$ or $\Psi^r_{\alpha}$ while
 $\Zvector_{(f)i}$ denotes
$\Qvector_{(f)i}$ or $\psivector_{(f) i \alpha}$.

\begin{figure}
\centerline{\epsfbox{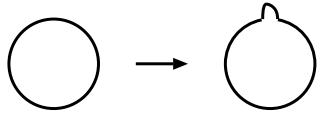}}
\caption{\small infinitesimal deformation of a closed string}
\label{closedkin}
\end{figure}

\begin{figure}
\vspace*{1cm}
\centerline{\epsfbox{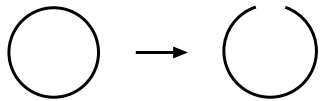}}
\caption{\small closed-open transition}
\label{closed-open}
\end{figure}

\begin{figure}
\vspace*{1cm} 
\centerline{\epsfbox{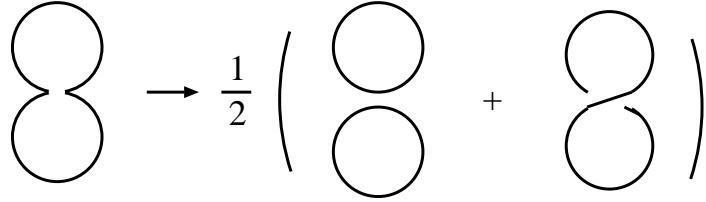}}
\caption{\small splitting and twisting of a closed string}
\label{closed}
\end{figure}

\begin{figure}
\vspace*{1cm} 
\centerline{\epsfbox{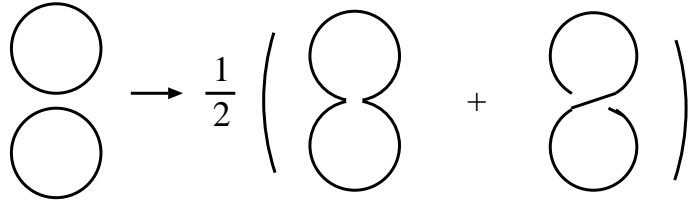}}
\caption{\small joining of two closed strings}
\label{closedclosed}
\end{figure}

\begin{figure}
\vspace*{1cm}
\centerline{\epsfbox{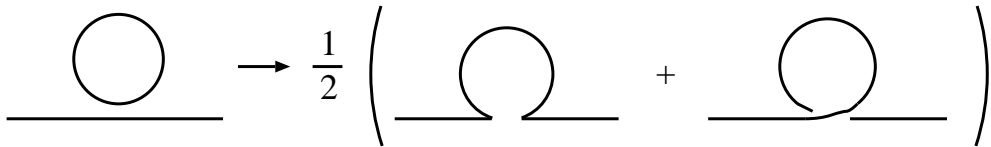}}
\caption{\small joining of a closed string and an open string}
\label{openclosed}
\end{figure}

We will exhibit eqs.
(\ref{loopeq1}) $\sim$ (\ref{loopeq3}) in the form of loop equations.
 We will repeatedly use
\beq
\sum_{r=1}^{2 k^2 \pm k} (T^r)_{i}^{\; j} (T^r)_{k}^{\; l}
=\frac{1}{2} (\delta_{i}^{\; l} \delta^{j}_{\; k}
\mp F^{-1}_{ik} F^{lj}) \;\;,
\eeq
which is nothing but the expression for the projector (\ref{projector}).
In these equations below, 
\beqn
P^{(i)}_{n}=\left\{ \begin{array}{l} 
                     p^{(i)M}_{n} \; \mbox{if} \; X^r=v^{r}_{M} \;, \\
                     -\bar{\eta}^{(i)}_{n}  \; \mbox{if} \; X^r=\Psi^r \;,
                    \end{array} \right.   \;\;\;
K^{(i)}_{n}=\left\{ \begin{array}{l} 
                     k^{(i)M}_{n} \; \mbox{if} \; X^r=v^{r}_{M} \;, \\
                     -\bar{\zeta}^{(i)}_{n}  \; \mbox{if} \; X^r=\Psi^r \;,
                    \end{array} \right. 
\eeqn               
and $\Lambda^{(i)}$ not multiplied by $\Pi$  represents either
 $\Xi^{(i)}$ or $\Theta^{(i)}$.
The symbol $\hat{b}$  denotes an omission of 
 the $b$-th closed or open loop.

\beqn
\bullet
(\ref{loopeq1}) \Rightarrow  \; 0  =
(1)\; \mbox{\underline{kinetic term (Fig. \ref{closedkin}),
 \ref{closed-open})}}
\;+\; (2)\; \mbox{\underline{splitting and twisting
 (Fig. \ref{closed})}} \\
+ (3)\; \mbox{\underline{joining with a closed string
 (Fig. \ref{closedclosed})}}
\;+\; (4)\; \mbox{\underline{joining with an open string
 (Fig. \ref{openclosed})}} \;\;.  \nonumber
\eeqn
Here
\beqn
 &(1)& = 
\frac{1}{g^2} \left\langle  \left( \delta_{X} \Phi[(1); X^r]  \right) 
\Phi[(2)] \cdots \Phi[(N)]  \Psi[(1)] \cdots \Psi[(L)] \right\rangle \;\;,
  \label{eq:loop1kin} \\
&(2)& =
\left\langle
\left( -\frac{i}{2}\sum_{n=n^{(1)}_{0}}^{n^{(1)}_{1}} P^{(1)}_{n}  \right)
 \left\{ \Phi[p^{(1)}_{.},\eta^{(1)}_{.};n^{(1)}_{1},n+1]
\Phi[p^{(1)}_{.},\eta^{(1)}_{.};n,n^{(1)}+1] \right.\right.\nonumber\\
 &\pm &   \left. Tr(U[p^{(1)}_{.},\eta^{(1)}_{.};n,n^{(1)}_1+1]
U[\mp p^{(1)}_{.},\mp \eta^{(1)}_{.};n+1,n^{(1)}_1]) \right\}
 \left. 
\Phi[(2)] \cdots \Phi[(N)]  \Psi[(1)] \cdots \Psi[(L)] \right\rangle
\nonumber\\
&+& \left\langle
 \left( -\frac{i}{2}\sum_{n=n^{(1)}_{1}+1}^{n^{(1)}_{2}} P^{(1)}_{n} \right)
 \left\{ \Phi[p^{(1)}_{.},\eta^{(1)}_{.};n,n^{(1)}_{1}+1]
\Phi[p^{(1)}_{.},\eta^{(1)}_{.};n^{(1)},n+1] \right.\right.\nonumber\\
 &\pm& \!\!\!\!\left. Tr(U[p^{(1)}_{.},\eta^{(1)}_{.};n^{(1)}_{1},n+1]
U[\mp p^{(1)}_{.},\mp \eta^{(1)}_{.};n^{(1)}_{1}+1,n]) \right\}
  \left. 
\Phi[(2)] \cdots \Phi[(N)]  \Psi[(1)] \cdots \Psi[(L)] \right\rangle ,
 \nonumber \\
  &&  \;\; \\
 &(3)& = 
\left\langle
\left( -\frac{i}{2}\sum_{b=2}^{N} \sum_{n=n^{(b)}_{0}}^{n^{(b)}_{1}}
 P^{(b)}_{n} \right)
 \left\{ Tr(U[p^{(1)}_{.},\eta^{(1)}_{.};n^{(1)}_{1},
n^{(1)}_{1}+1] U[p^{(b)}_{.},\eta^{(b)}_{.};n,n+1]) \right.\right.\nonumber\\
&& \mp \left. Tr(U[\mp p^{(1)}_{.},\mp
 \eta^{(1)}_{.};n^{(1)}_{1}+1,n^{(1)}_{1}]
U[p^{(b)}_{.},\eta^{(b)}_{.};n,n+1]) \right\}  \nonumber\\
&&\left. \Phi[(2)] \cdots \hat{b} \cdots \Phi[(N)]  
\Psi[(1)] \cdots \Psi[(L)] \right\rangle \;\;, \\
 &(4)& =
\left\langle 
\left( -\frac{i}{2}\sum_{b=1}^{L} \sum_{l=l^{(b)}_{0}}^{l^{(b)}_{1}}
 K^{(b)}_{l} \right) \right. \nonumber \\
  && \left\{  \left( \Lambda^{(b)'} \Pivector_{(f^{(b)'})} \right) F
U[k^{(b)}_{.},\zeta^{(b)}_{.};l^{(b)}_{1},l+1]
U[p^{(1)}_{.},\eta^{(1)}_{.};n^{(1)}_{1},n^{(1)}_{1}+1]
U[k^{(b)}_{.}, \zeta^{(b)}_{.};l,l^{(b)}_{0}]
 \left(  \Lambda^{(b)}  \Pivector_{(f^{(b)})} \right) \right. \nonumber\\
 &\mp& \!\!\!\!\left. \left( \Lambda^{(b)'}  \Pivector_{(f^{(b)'})}  \right) F
U[k^{(b)}_{.},\zeta^{(b)}_{.};l^{(b)}_{1},l+1]
U[\mp p^{(1)}_{.},\mp \eta^{(1)}_{.};n^{(1)}_{1}+1,n^{(1)}_{1}]
U[k^{(b)}_{.},\zeta^{(b)}_{.};l,l^{(b)}_{0}]
\left(  \Lambda^{(b)} \Pivector_{(f^{(b)})} \right) \right\} \nonumber\\
&& \hspace*{4cm}\left. 
\Phi[(2)] \cdots \Phi[(N)]  
\Psi[(1)] \cdots \hat{b} \cdots \Psi[(L)] \right\rangle \;\;\;.
\label{loopeq1complete}
\eeqn
  The term  $\delta_{X} \Phi[(1); X^r]$
    comes from the variation of the action and
 contains terms representing  closed-open transition.
 For its explicit form, see the original paper \cite{Itsu}.
  We present the pictures associated with the terms
  $(1) \sim (4)$  as figures.

 

\begin{figure}
\vspace*{1cm} 
\centerline{\epsfbox{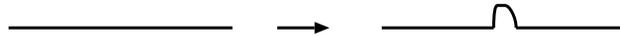}}
\caption{\small infinitesimal deformation of an open string: case one}
\label{openkin1}
\end{figure}

\begin{figure}
\vspace*{1cm} 
\centerline{\epsfbox{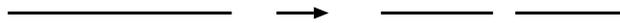}}
\caption{\small splitting of an open string}
\label{open2}
\end{figure}

\begin{figure}
\vspace*{1cm} 
\centerline{\epsfbox{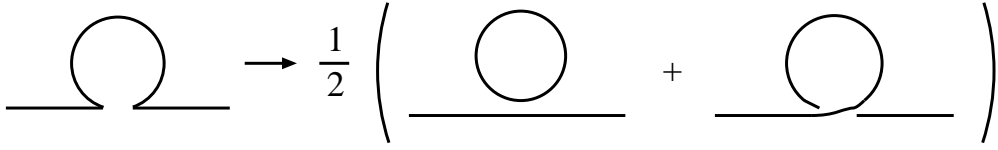}}
\caption{\small splitting and twisting of an open string}
\label{open1}
\end{figure}

\begin{figure}
\vspace*{1cm} 
\centerline{\epsfbox{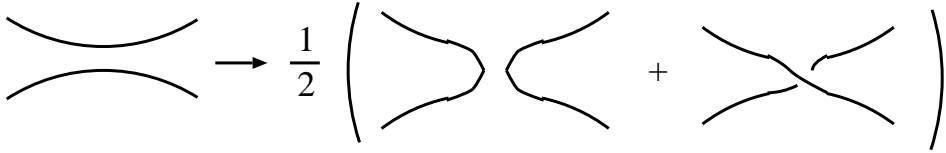}}
\caption{\small joining of two open strings: case one}
\label{openopen1}
\end{figure}

\beqn
\bullet
(\ref{loopeq2}) &\Rightarrow&  \;\;0 =
(1)\; \mbox{\underline{kinetic term (Fig. \ref{openkin1},\ref{open2})}}
 + (2)\; \mbox{\underline{splitting and twisting (Fig.\ref{open1} )}} \\
&+& (3)\;\mbox{\underline{joining with a closed string
 (Fig. \ref{openclosed})}}
+(4)\;\mbox{\underline{joining with an open string (Fig. \ref{openopen1})}}
\;\;.
\nonumber 
\eeqn
 We will  present here only the figures associated with
  $(1) \sim (4)$ of eq. (\ref{loopeq2}).


\begin{figure}
\vspace*{1cm} 
\centerline{\epsfbox{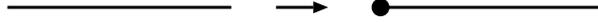}}
\caption{\small infinitesimal deformation of an open string: case two}
\label{openkin2}
\end{figure}

\begin{figure}
\vspace*{1cm} 
\centerline{\epsfbox{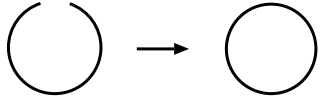}}
\caption{\small open-closed transition}
\label{open-closed}
\end{figure}

\begin{figure}
\vspace*{1cm} 
\centerline{\epsfbox{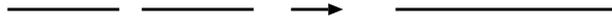}}
\caption{\small joining of two open strings: case two}
\label{openopen2}
\end{figure}

\beqn
\bullet
(\ref{loopeq3}) &\Rightarrow&  \;\;0  =
(1)\; \mbox{\underline{kinetic term (Fig. \ref{openkin2})}} 
 + (2)\; \mbox{\underline{open-closed transition 
(Fig. \ref{open-closed})}}      \nonumber \\
&+& (3)\;
 \mbox{\underline{joining with an open string (Fig. \ref{openopen2})}} \;\;.
\eeqn
 Again, we will present here only the figures associated with
  $(1) \sim (3)$ of eq. (\ref{loopeq3}).


\begin{figure}
\vspace*{1cm} 
\centerline{\epsfbox{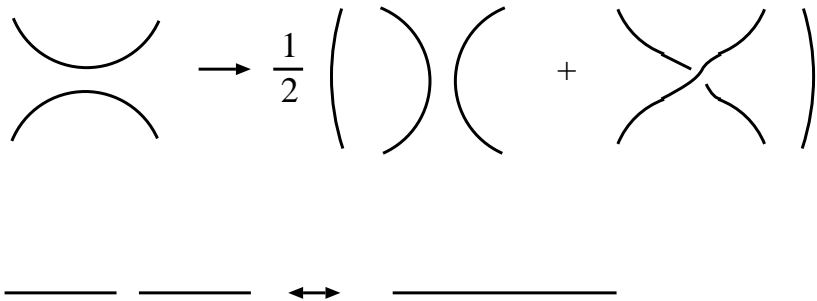}}
\caption{\small two kinds of elementary local processes}
\label{elementaryprocess}
\end{figure}

  We have checked that all  three of the loop equations
 are expressed by the closed and open loops
 $\Phi$ and $\Psi$ and their derivatives with respect to the sources
 introduced.
For example, the expression
$\psivector^{\ast} \cdot  \bar{\sigma}^m
U[p^{(1)}_{.},\eta^{(1)}_{.};n^{(1)}_{1},n^{(1)}_{1}+1] \psivector$
  contained in  $\left( \delta_{X} \Phi[(1); X^r] \right)$
in  eq. (\ref{eq:loop1kin}) is represented as
\beq 
\sum_{f=1}^{2n_f}\bar{\sigma}^{\dot{\alpha}\alpha m} 
\frac{\partial}{\partial\bar{\theta}^{'\dot{\alpha}}}
\frac{\partial}{\partial\theta^{\alpha}} \Psi_{f f}[p^{(1)}_{.},
\eta^{(1)};n^{(1)}_{1},n^{(1)}_{1}+1;\Lambda',\Lambda]\;\;.
\eeq
In this sense, the set of loop equations we have derived is closed. 
 It is noteworthy that all of the terms in the above loop equations
are either an infinitesimal deformation of a loop or a consequence from 
  the two elementary local processes of loops which are illustrated 
   in Fig. \ref{elementaryprocess}.

 It is interesting to discuss the system of loop equations we have derived 
in the light of string field theory.  In addition to the lightcone
superstring field theory constructed earlier in \cite{GSSFT}, there is 
now gauge invariant string field theory for closed-open bosonic system
\cite{KT}. We find that the types of the interaction terms of our equations
are in complete agreement with the interaction vertices seen in 
\cite{GSSFT}  and
the second paper of \cite{KT}. In particular, Figures $2\sim5, 7\sim9,
11\sim12$  for the interactions of our equations are in accordance with
  $U, V_{\infty}, V_{3}^{c}, U_{\Omega}, V_{3}^{0}, V_{\alpha}, V_{4}^{0}$
  of \cite{KT}. 
   While BRS invariance  determines  the coefficients of the interaction
 vertices in  \cite{KT}, the (bare) coefficients are already determined
 in our case  from the first quantized action.  This may give us insight
 into  properties of the model which are not revealed.

\subsection{Linearized loop equations and  a free string}
  Let us consider the all three loop equations
eqs. (\ref{loopeq1}), ({\ref{loopeq2})  and (\ref{loopeq3}) in the linearized
 approximation,  namely, ignoring  the joining and splitting of the loops. 
  Let us first introduce a variable conjugate to $p_{A n}$ or $k_{A n}$
  and that to $\eta_{n}$ or $\zeta_{n}$  by
\beqn
\hat{X}^{A}_{n} &=& i \frac{\delta}{\delta p_{A n}} \; \mbox{or} \;
                 \; i \frac{\delta}{\delta k_{A n}}  \;\;, \\
\hat{\Psi}_{n} &=&  i \frac{\delta}{\delta \eta_{n}} \; \mbox{or} \;
               \;  i \frac{\delta}{\delta \zeta_{n}}\;\;.
\eeqn
By acting $\hat{X}^{A}_{n}$ and $\hat{\Psi}_{n}$ on a loop, we obtain
   respectively  an  operator insertion
of $v^{A}$ and that of $\Psi$ at point $n$  on the loop.

Now consider   eq. (\ref{loopeq1}) and  eq. (\ref{loopeq2}) 
for  the case $X^r=v_{M}^{r}$,   multiplying them by $p_{nM}^{(1)}$ and
$k_{nM}^{(1)}$ respectively.     Consistency requires that,
 for these terms, we must take into account
 the term from the interactions which represents splitting of 
  a loop  with infinitesimal length.
  This in fact occurs when
 the splitting point $n$ coincides with the point $n_{1}^{(1)}$ 
at which $T^r$ is inserted.
We obtain
\beqn
0&=&\frac{1}{g^2} p_{nM}^{(1)}
\left\langle  \left( \delta_{X} \Phi[(1); v_{M}^{r}]  \right) 
\Phi[(2)] \cdots \Phi[(N)]  \Psi[(1)] \cdots \Psi[(L)]
 \right\rangle \nonumber\\
&&-\frac{i}{2}2k p_{n}^{(1)2}
\left\langle  \Phi[(1)] 
\Phi[(2)] \cdots \Phi[(N)]  \Psi[(1)] \cdots \Psi[(L)] \right\rangle  \;\;,\\
0&=&\frac{1}{g^2} k_{nM}^{(1)}
\left\langle  \left( \delta_{X} \Psi[(1); v_{M}^{r}]  \right) 
\Phi[(1)] \cdots \Phi[(N)]  \Psi[(2)] \cdots \Psi[(L)] \right\rangle \nonumber\\
&&-\frac{i}{2}2k k_{n}^{(1)2}
\left\langle  \Phi[(1)] 
\Phi[(2)] \cdots \Phi[(N)]  \Psi[(1)] \cdots \Psi[(L)] \right\rangle \;\;.
\eeqn
These equations lead to the half of the Virasoro conditions \cite{FKKT}:
\beqn
0&=&(p_{n}^{(1)2}+\hat{X}_{n}^{(1)'2}+\mbox{(fermionic part)}) \nonumber\\
&&\left\langle  \Phi[(1)] 
\Phi[(2)] \cdots \Phi[(N)]  \Psi[(1)] \cdots \Psi[(L)] \right\rangle  \;\;, \\
0&=&(k_{n}^{(1)2}+\hat{X}_{n}^{(1)'2}+\mbox{(fermionic part)}) \nonumber\\
&&\left\langle  \Phi[(1)] 
\Phi[(2)] \cdots \Phi[(N)]  \Psi[(1)] \cdots \Psi[(L)] \right\rangle \;\;, 
\eeqn
where $\prime$  implies  taking  a difference between  two adjacent
 points $n$  and $n+1$.
The reparametrization invariance of the Wilson loops leads to
the remaining half of the Virasoro conditions:
\beqn
0&=&(p_{n}^{(1)M}\hat{X}_{nM}^{(1)'}+\mbox{(fermionic part)}) \nonumber\\
&&\left\langle  \Phi[(1)] 
\Phi[(2)] \cdots \Phi[(N)]  \Psi[(1)] \cdots \Psi[(L)] \right\rangle \;\;, \\
0&=&(k_{n}^{(1)M}\hat{X}_{nM}^{(1)'}+\mbox{(fermionic part)}) \nonumber\\
&&\left\langle  \Phi[(1)] 
\Phi[(2)] \cdots \Phi[(N)]  \Psi[(1)] \cdots \Psi[(L)] \right\rangle \;\;.
\eeqn

Next, let us consider  eq. (\ref{loopeq3}), ignoring joining and splitting
 of the loops.  Again consistency appears to require that
 we drop the cubic terms consisting of $\Qvector$ and $\Qvector^{\ast}$
   in   $\delta_{\Zvector} \Psi[(1);\Zvector]$ .
  To write explicitly, the following expression   must vanish
\beqn
&& \left\{ \Qvector^{\ast}_{(f)}(v_{\nu}v^{\nu}+[\Phi_I,\Phi^I])
+(\Qvector \Sigma)_{(f)}F[\Phi_{2}^{\dagger},\Phi_{3}^{\dagger}]
+(\Qvector^{\ast}M^2)_{(f)}+2 (\Qvector^{\ast}M)_{(f)}v_4
- i \sqrt{2} \psivector^{\ast}_{(f)}\bar{\lambda}   \right.  \nonumber\\
&& \left. -\sqrt{2}(\psivector \Sigma)_{(f)} F \psi_{\Phi_1} \right\} 
U[k^{(1)}_{.},\zeta^{(1)}_{.};l^{(1)}_{1},l^{(1)}_{0}] 
(\Lambda^{(1)} \Pivector_{f^{(1)}}) \approx  0 
  \label{linearizedloopeq1} \;\;\;, \\
&& \left\{ \psivector^{\ast}_{(f)} \bar{\sigma}^{m} v_{m}
+i \sqrt{2} \Qvector_{(f)}^{\ast} \lambda 
+ \left( \psivector  \Sigma F (\sqrt{2}\Phi_1 + M) \right)_{(f)}  \right.
   \nonumber \\
&&\left. +\sqrt{2} (\Qvector  \Sigma F \psi_{\Phi_1})_{(f)} \right\}
U[k^{(1)}_{.},\zeta^{(1)}_{.};l^{(1)}_{1},l^{(1)}_{0}] 
(\Lambda^{(1)} \Pivector_{f^{(1)}}) \approx  0   \;\;\;,
\label{linearizedloopeq2}
\eeqn
when inserted in 
\beq
\left\langle  \Phi[(1)] 
\Phi[(2)] \cdots \Phi[(N)] \Psi[(2)] \cdots \Psi[(L)]  \right\rangle \;\;.
\eeq

  As we stated before, the lefthand sides of eqs.
(\ref{linearizedloopeq1}) and (\ref{linearizedloopeq2}) are expressible
as an open loop with some functions of $\hat{X}^{A}_{l^{(1)}_{1}}$  and
$\hat{\Psi}_{l^{(1)}_{1}}$ acting on the loop. Let us see by inspection how
eqs. (\ref{linearizedloopeq1}) and (\ref{linearizedloopeq2}) are satisfied by
 the source functions alone.
Consider  the following configuration of $\hat{X}_{n}$ and $\hat{\Psi}_{n}\;,$
$n= l^{(1)}_{1}$:
\beqn
  && \hat{X}^{\mu}    \approx 0  \;\; \mbox{for}\;\; \mu = 0,1,2,3,7 \;\;,
  \;\;\;\hat{X}^{4}=\pm m_f   \;\;\;.  \label{boundary1}   \\
&& \hat{X}^{'I} \approx 0  \;\; \mbox{for}\;\;  I = 5,6,8,9 
 \label{boundary2}  \\
  &&   \hat{\Gamma}_{3} \hat{\Psi}  \approx  -\hat{\Psi} \;\;\;,
\label{boundary3}
\eeqn
where $\hat{\Gamma}_{3}  \equiv \Gamma_5 \Gamma_6 \Gamma_8 \Gamma_9$.  
Again  these equations  should be understood in the sense of 
   an insertion at the end point of the open loop.

  Eqs. (\ref{boundary1}), (\ref{boundary2}), and (\ref{boundary3}) tell us
 that the open loop $\Psi[(1)]$   obeys the Dirichlet boundary
conditions for $0,1,2,3,4,7$ directions and the Neumann boundary 
conditions for $5,6,8,9$ directions.

 We find that  the configuration given by eqs. (\ref{boundary1}),
(\ref{boundary2}) and (\ref{boundary3}) solves the linearized loop equations
(\ref{linearizedloopeq1}), (\ref{linearizedloopeq2}).
This configuration clearly tells us the existence of $n_f$ $D3$ branes
and their mirrors  each of which is  at a distance $\pm m_f$ away from the
 orientifold surface in the fourth direction.
  This conclusion  consolidates both the semiclassical picture
 in section three and the picture emerging from the
 fermionic integrations in section four.

\section*{Acknowledgements}
We thank the organizers of the Nishinomiya symposium and
  the YITP workshop. This work is supported in part
 by the Grant-in-Aid for Scientific Research (10640268,10740121)
 and by the Grant-in-Aid  for Scientific Research Fund (97319)
from the Ministry of Education, Science and Culture, Japan.


\begin{thebibliography}{99}


 

\bibitem{IT1}
H. Itoyama and A. Tokura, {\sl Prog. Theor. Phys.} {\bf 99} (1998)129,
hep-th/9708123.

\bibitem{IT2}
H. Itoyama and A. Tokura, {\sl Phys. Rev.} {\bf D58} (1998) 026002, 
hep-th/9801084.

\bibitem{IKKT}
N. Ishibashi, H. Kawai, Y. Kitazawa and A. Tsuchiya,
 {\sl Nucl. Phys.} {\bf B498} (1997)467, hep-th/9612115. 


\bibitem{GSano} 
M. B. Green and J. H. Schwarz, {\sl Phys. Lett.} {\bf 149B} (1984) 117.

\bibitem{Schild}
A. Schild,
{\sl Phys. Rev.} {\bf D16} (1977)1722.

\bibitem{BFSS}
T. Banks, W. Fischler, S. Shenker and L. Susskind,
{\sl Phys. Rev.} {\bf D55 }(1997) 5112, hep-th/9610043.


\bibitem{M}
 E. Witten, {\sl Nucl. Phys.} {\bf B443} (1995)85.
 
\bibitem{Itsu}
  H. Itoyama and A. Tsuchiya  hep-th/9812177.

\bibitem{IM}
  H. Itoyama and T. Matsuo, {\sl Phys.Lett.} {\bf B439} (1998) 46,
 hep-th/9806139.

\bibitem{CIK}
  B. Chen, H. Itoyama and H. Kihara, hep-th/9810237.


\bibitem{Mtwist}
U.H. Danielsson and G. Ferretti,
{\sl Int. J. Mod. Phys.} {\bf A12} (1997)4581; 
L. Motl,
preprint hep-th/9612198; 
N. Kim and S.J. Rey,
{\sl Nucl.Phys.} {\bf B504} (1997)189.


\bibitem{Ba}
S. Kachru and E. Silverstein, {\sl Phys. Lett.} {\bf 396B} (1997)70;
D. A. Lowe, {\sl Nucl.Phys.} {\bf B501} (1977) 134,
{\sl Phys. Lett.} {\bf 403B} (1997)243;
T. Banks, N. Seiberg and E. Silverstein,
{\sl Phys. Lett.} {\bf 401B} (1997)30.


\bibitem{GS}
M. B. Green and J. H. Schwarz, 
{\sl Phys. Lett.} {\bf B151} (1985)21. 

\bibitem{IMo}
 H. Itoyama and P. Moxhay, {\sl Nucl. Phys.} {\bf B293} (1987)685.


\bibitem{AIKKT}
H. Aoki, S. Iso, H. Kawai, Y. Kitazawa and
T. Tada, hep-th/9802085.

\bibitem{HNT}
T. Hotta, J. Nishimura and A. Tsuchiya, hep-th/9811220


\bibitem{Pol} 
J. Polchinski,
{\sl Phys. Rev. Lett.} {\bf 75} (1995)4724.
\\
J. Polchinski, {\it TASI Lectures on D-branes \/} 
hep-th/9611050.

\bibitem{WT}
W. Taylor IV, {\sl Phys. Lett.} {\bf 394B} (1997) 283  
hep-th/9611042; O.J. Ganor, S. Ramgoolam and W. Taylor IV,
{\sl Nucl. Phys.} {\bf B492} (1997) 191 hep-th/9611202.

\bibitem{AIKT}
 Y. Arakane, H. Itoyama, H. Kunitomo and A. Tokura, 
{\sl Nucl. Phys.} {\bf B486} (1997)149.

\bibitem{Seiberg}
N. Seiberg, hep-th/9608011.

\bibitem{PW}
 J. Polchinski and E. Witten, {\sl Nucl. Phys.} {\bf B460} (1996)525, 
hep-th/9510169.

\bibitem{AI}
Y. Arakane and H. Itoyama, in preparation

\bibitem{V}
C. Vafa, {\sl Nucl.Phys.} {\bf B469} (1996) 403, hep-th/9602022.

\bibitem{Sen}
A. Sen, {\sl Nucl.Phys.} {\bf B475} (1996) 562, hep-th/9605150.



\bibitem{Nepomechie}
  R. I. Nepomechie, Phys. Rev. {D31} (1985) 1921.


\bibitem{GSSFT}
M.B. Green, J. H. Schwarz {\sl Nucl. Phys.} {\bf B243} (1984) 475 .



\bibitem{KT}
  T. Kugo and T. Takahashi, {\sl Prog.Theor.Phys.} {\bf 99} (1998) 649,
 hep-th/9711100 ; T. Asakawa, T. Kugo and T. Takahashi, hep-th/9807066.



\bibitem{BSS}
L. Brink, J. Scherk and J. H. Schwarz,
{\sl Nucl.Phys.} {\bf B121} (1977) 77.


\bibitem{FKKT}
M. Fukuma, H. Kawai, Y. Kitazawa and A. Tsuchiya,
{\sl Nucl. Phys.} {\bf B510} (1998)158, hep-th/9705128. 


\end{thebibliography}
\end{document}